\documentclass[12pt,twocolumn,twoside]{article}

\usepackage[para]{threeparttablex}
\newcommand{\mtnote}[1]{\textsuperscript{\TPTtagStyle{#1}}}
\usepackage{multirow}
\usepackage{multicol}
\usepackage{wrapfig}
\usepackage{titlesec}
\usepackage{blkarray}
\usepackage{fancyhdr}
\usepackage{kbordermatrix}
\usepackage[numbers,comma]{natbib}
\usepackage{graphicx, caption}
\usepackage{subfig}
\usepackage{enumitem,booktabs}
\usepackage{rotating}
\usepackage{amsmath, amssymb, amsthm}
\usepackage{algorithm}
\usepackage[noend]{algpseudocode}
\usepackage{tikz}
\usepackage{adjustbox}
\usepackage{multirow}
\usetikzlibrary{patterns}
\usepackage{pgfplots}
\tikzset{
    axis break gap/.initial=1mm
}
\usepackage{multirow}
\usepackage{siunitx}
\usepackage{booktabs, longtable}
\usepackage[top=1in,bottom=1in,left=1in,right=1in]{geometry}
\usepackage{fancyhdr}
\usepackage{hyperref}
\usepackage{color}
\usepackage{setspace,verbatim}
\usepackage{authblk}
\usepackage{transparent}

\usepackage{xcolor,colortbl}
\usepackage{url}
\usepackage{hyperref}

\usepackage{graphicx}

\definecolor{Gray}{gray}{0.85}
\definecolor{LightCyan}{rgb}{0.88,1,1}
\usepackage{array}
\newcolumntype{a}{>{\columncolor{Gray}}c}
\newcolumntype{b}{>{\columncolor{white}}c}
\newcolumntype{C}[1]{>{\centering\let\newline\\\arraybackslash\hspace{0pt}}m{#1}}
\newcolumntype{L}[1]{>{\raggedright\let\newline\\\arraybackslash\hspace{0pt}}m{#1}}

\usepackage[customcolors]{hf-tikz}

\tikzset{style green/.style={
    set fill color=green!50!lime!60,
    set border color=white,
  },
  style cyan/.style={
    set fill color=cyan!90!blue!60,
    set border color=white,
  },
  style orange/.style={
    set fill color=orange!80!red!60,
    set border color=white,
  },
  hor/.style={
    above left offset={-0.15,0.31},
    below right offset={0.15,-0.125},
    #1
  },
  ver/.style={
    above left offset={-0.1,0.3},
    below right offset={0.15,-0.15},
    #1
  }
}

\titleformat*{\section}{\normalsize\bfseries\sffamily}
\titleformat*{\subsection}{\small\bfseries\sffamily}
\titleformat*{\subsubsection}{\small\bfseries\sffamily}

\usepackage[font=small,labelfont=bf]{caption}

\newcommand\shorttitle{CAMD Review}
\newcommand\authors{Austin, Sahinidis, and Trahan}

\fancypagestyle{firststyle}
{
   \fancyhf{}

   \fancyfoot[L]{\footnotesize\scshape \copyright 2017. This manuscript version is made available under the CC-BY-NC-ND 4.0 license \url{http://creativecommons.org/licenses/by-nc-nd/4.0/}.  The formal publication of this article is available at \url{http://dx.doi.org/10.1016/j.cherd.2016.10.014}.}
}

\title{\textbf{Computer-aided molecular design: An introduction and review of tools, applications, and solution techniques}
}
\author[1]{Nick~D.~Austin}
\author[1]{Nikolaos~V.~Sahinidis}
\author[2]{Daniel~W.~Trahan}
\affil[1]{Carnegie Mellon University, Pittsburgh, PA, USA}
\affil[2]{The Dow Chemical Company, Freeport, TX, USA}
\date{}

\begin{document}

\begin{singlespacing}
\twocolumn[{%
\maketitle
\thispagestyle{firststyle}

\bibliographystyle{plainnat}


\footnotesize
\vspace{-1cm}
\begin{abstract}
This article provides an introduction to and review of the field of computer-aided molecular design (CAMD). It is intended to be approachable for the absolute beginner as well as useful to the seasoned CAMD practitioner.  We begin by discussing various quantitative structure-property relationships (QSPRs) which have been demonstrated to work well with CAMD problems.  The methods discussed in this article are (1) group contribution methods, (2) topological indices, and (3) signature descriptors.  Next, we present general optimization formulations for various forms of the CAMD problem.  Common design constraints are discussed and structural feasibility constraints are provided for the three types of QSPRs addressed.  We then detail useful techniques for approaching CAMD optimization problems, including decomposition methods, heuristic approaches, and mathematical programming strategies.  Finally, we discuss many applications that have been addressed using CAMD.
\end{abstract}


\vspace*{0.8cm}

}]

\footnotesize

\section{Introduction}
\label{intro}
The application of chemistry to manipulate the natural world has its earliest examples in metalworking and pottery~\cite{history_of_chemistry}, with some pottery artifacts reported to be as old as 20,000 years~\cite{wzg+12}.  Chemical products have since played an important role in history and left an indelible mark on the way we live and work.  Early history included basic incendiary fuels, perfumes, and soap as some of the first widespread uses of chemicals.  The modern age has witnessed an unprecedented expansion of chemical products, including pesticides, fuels for transportation and electricity, pharmaceuticals, plastics, and a broad array of industrial and consumer products.  True to these historical trends, few things are as pervasive as chemical products in 21st century life, and there are an ever-increasing number of new chemical applications which require specialized compounds.

The process of determining new and suitable chemicals for a certain application can be generally termed chemical product design~\cite{cm11}.  Chemical product design has long been a laborious, trial-and-error procedure, limited often by a fixed amount of chemical, time, and financial resources.  Design efforts are often high-throughput and tend to focus on a small class of compounds or structural analogues of known chemicals.  Accordingly, the so-called ``design space''---the set of unique molecular structures considered---is often quite small for these product design problems, especially considering the massive design space of all possible chemical structures.  It is then clear that to keep pace with the growing demand for new chemical products and to adequately explore the full chemical design space, other approaches must be considered.  Fortunately, the availability and efficiency of computational resources makes these design problems more tenable than ever before.  Noteworthy among computational approaches is the field of computer-aided molecular design (CAMD), which leverages the simplicity of semi-empirical quantitative structure-property relationships (QSPRs) in conjunction with fast and efficient numerical optimization algorithms.

CAMD has its roots in the 1980's, although the general use of computers in chemistry pre-dates this by a few decades.  Stated formally, the CAMD problem concerns designing an optimal molecular structure(s) for a certain application.  CAMD combines molecular modeling techniques, thermodynamics, and numerical optimization to design good or optimal molecular structures, many of them often completely novel.  Advances in chemical modeling in the last few decades have greatly benefited CAMD, and practitioners are now capable of relating chemical structures to properties at several levels of accuracy (molecular mechanics, semi-empirical, \textit{ab initio}).  Though CAMD often uses semi-empirical modeling techniques for their simplicity and efficiency, new approaches incorporating more accurate methods are emerging.  Modern combinatorial optimization techniques are also essential for CAMD, enabling the optimization over staggeringly large design spaces which would otherwise be inaccessible (using enumeration algorithms, for example).

This article is intended to provide a description of popular QSPRs used in CAMD, the CAMD problem itself, and several solution approaches to its various forms.  There exist several CAMD reviews~\cite{gani_camd_review,ng_camd_review} that provide a thorough coverage of many relevant applications and milestones of CAMD and are good resources for the interested reader.  While also providing a review of CAMD literature, the current article first motivates the field from the ground up. In this way, this review is intended to be accessible to readers of all levels as it presents many essential CAMD tools and techniques in detail.  We also provide a discussion of various techniques and a review of CAMD applications, so this document will also be of interest to the more advanced CAMD practitioner. We begin in Section~\ref{camd_qsprs} by detailing three popular classes of QSPRs which are often used in CAMD: (1) group contribution methods; (2) topological indices; and (3) signature descriptors.   Next, in Section~\ref{camd_optimization_prob}, we present the CAMD problem from a mathematical programming perspective, discussing various classes of the single-molecule design problem as well as CAMD problems considering mixtures of molecules and those involving the simultaneous design of a chemical product and the process it is a part of.  In this section, several other important design considerations are presented, including a few important constraints to ensure practical solutions as well as the chemical feasibility of the designed structures.  In Section~\ref{camd_techniques}, various solution techniques for the CAMD problem are discussed, including mathematical optimization strategies, decomposition methods, and heuristic approaches.  Finally, in Section~\ref{camd_applications}, we provide a diverse though non-exhaustive review of applications of CAMD problems.

\section{Popular types of QSPRs in CAMD}
\label{camd_qsprs}

The CAMD problem attempts to choose optimal (or simply good) molecules for some purpose from the space of theoretically possible chemical structures.  At first glance, the CAMD problem must consider a very abstract chemical design space of atoms, bonds, aromaticity, structural isomers, electronic effects, etc.  Though many of these features are certainly what gives molecules their specific properties and chemical functionality, they are difficult to build into any type of optimization scheme.  This is primarily because there is no available immediate relationship between an arbitrary chemical structure and its performance or suitability regarding a specific application.  In order to ``rank'' different structures and choose an optimal one, we must have some efficient way to quantify the properties and performance of each structure.

A second issue is the sheer size of the chemical search space.  At the time of writing this article, the CAS registry~\cite{cas} reports over 115 million unique organic and inorganic structures.  This number only represents compounds which have been synthesized and cataloged, and it is already far too large for every structure to be considered in any type of trial-and-error design scheme.  This number is also only a fraction of the theoretically possible chemical space, which some estimates indicate may contain more than $10^{60}$ unique molecules for small, drug-like structures~\cite{bcg96}.  Even with very efficient ways to estimate the performance of a certain structure, screening these structures using an enumeration strategy is far beyond current computational capacity.  For this reason, we also need to relate the chemical space to a space that can be utilized for combinatorial optimization, allowing us to design over the massive search space far more efficiently.

CAMD practitioners have relied on semi-empirical quantitative structure property relationships (QSPRs) to address both of these issues.  First, many semi-empirical methods delineate a clear connection between the abstract chemical space and a more practical space of quantitative properties.  These methods are also often simple and can be applied to estimate properties very efficiently.  Second, many of these methods break molecular structure into sub-molecular collections of atoms and bonds.  These molecular sub-structures are assumed to dictate a molecule's properties.  Using these types of representations of the molecular space, combinatorial optimization can be directly applied to molecular design problems.

\subsection{Group-contribution methods}

\begin{figure}
\begin{center}
\caption{Propanol represented by its groups\label{propanol_groups}}
\includegraphics[scale=1]{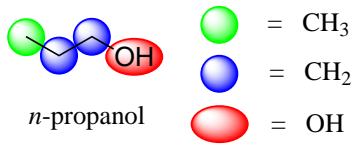}
\end{center}
\end{figure}

The most commonly used QSPRs in CAMD are group contribution (GC) methods.  These work under the assumption that a molecule's properties can be predicted by the number of occurrences of various molecular sub-structures called ``groups.''  For example, we may think to represent the simple molecule propanol as a combination of the groups $-\text{CH}_3$, $-\text{CH}_2-$, and $-\text{OH}$.  In this case, the dashes $(-)$ represent bonds to other groups.  In its group representation, propanol would no longer be thought of as the connected alcohol molecule, but rather as some collection of its constituent groups.  The group representation of propanol is shown in Fig.~\ref{propanol_groups}.

\begin{figure*}
\caption{Example usage of group contribution methods\label{gc_example}}
\footnotesize
\centering
     \begin{tabular}{ c c c }
     \centering
     \textbf{Original structure} & \textbf{Group representation}& \textbf{Number of occurrences} \\
     \toprule
     \multicolumn{2}{c}{
     \begin{tabular}{ c c }
     \includegraphics[scale=1.0]{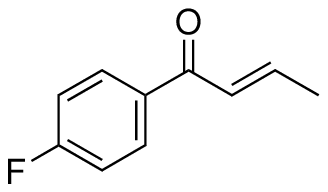}  &
     \includegraphics[scale=1.0]{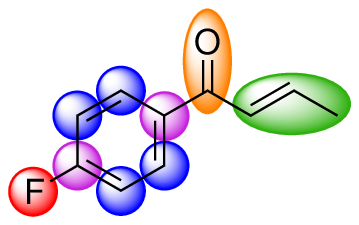}  \\
      \multicolumn{2}{c}{
	\colorbox{blue!15!white}{ 	
	  \begin{tabular}{ C{2.5cm} c }
	  \multicolumn{2}{c}{
	  \textbf{Estimating properties}} \\
	  \hline
	  Group composition vector &
	  $\begin{aligned}
	  n=&[0,\hdots,1,\hdots,2,\hdots, \\
	  & 4,\hdots,1,\hdots,1,\hdots]
	  \end{aligned}$
	   \vspace{2mm}\\	  	
	  Coefficient vector (example) &
	  $\begin{aligned}
	  c = & [\hdots,3.2,\hdots,-2.4,\hdots, \\
	  & 0.6,\hdots,1.2,\hdots,2.3,\hdots]
	  \end{aligned}$
	   \vspace{2mm} \\	
	  Property estimate &
	  $
	  \begin{aligned}
	  P = \sum\limits_g c_g &n_g \\
	   = 3.2(1)-2.4(2)+0.6&(4)+1.2(1)+2.3(1)
	  \end{aligned}$
	  \end{tabular}	
    }}
     \end{tabular}}
    &
    \hspace{-5mm}
    \begin{tabular}{ c C{0.6cm} c }
    \vspace{3mm} \adjustbox{valign=c}{\includegraphics[scale=1.0]{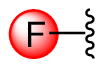}} && 1  \\
    \vspace{3mm} \adjustbox{valign=c}{\includegraphics[scale=1.0]{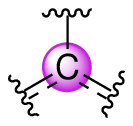}} && 2  \\
    \vspace{3mm}\adjustbox{valign=c}{\includegraphics[scale=1.0]{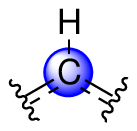}} && 4 \\
 	\vspace{3mm}\adjustbox{valign=c}{\includegraphics[scale=1.0]{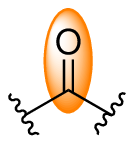}} && 1 \\
 	\vspace{3mm}\adjustbox{valign=c}{\includegraphics[scale=1.0]{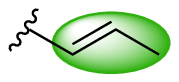}} &&1 \\
    \end{tabular}
\end{tabular}
\end{figure*}

Being QSPRs, group contribution methods translate the group representation of a molecular structure into an estimate for some property $P$.  To do this, group contribution methods define a vector $n$ that represents the number of occurrences of each of the groups.  Assuming we only have the three groups shown above, propanol's $n$ vector would look like $n=[1,2,1]$, where the entries in this vector represent the number of occurrences of the groups $-\text{CH}_3$, $-\text{CH}_2-$, and $-\text{OH}$, respectively.  Each of these groups $g$ would also be associated with a coefficient $c_g$ which quantifies its affect or ``contribution'' to a particular property $P$.  Properties are calculated as follows:
\begin{equation}
P=\sum\limits_g c_g n_g
\end{equation}
The vector of coefficients $c$ comes from regression of the property $P$ over a large dataset of different molecules.  To regress these parameters, the identity of all of the groups must be specified \textit{a priori}.  Returning to our example, one can easily imagine different sets of groups being used to describe propanol.  For example, the groups $-\text{CH}_3$, $-\text{CH}_2-$, and $-\text{CH}_2\text{OH}$ also completely account for the atoms in propanol, and these may provide a better fit for the regression problem.  For this reason, different group contribution methods to estimate different properties usually do not have consistent sets of groups.  Finally, we note that the vector $n$ is generally much bigger as many group contribution methods contain 50-100 groups.  Many group contribution methods make the additional assumption that groups cannot overlap, which means that $n$ is typically a sparse vector.  A pictorial example of the usage of group contribution methods is given in Fig.~\ref{gc_example}.  In this example, we apply a hypothetical GC method to a hypothetical molecule.  We show how a molecular structure is decomposed into its constituent groups and provide a count of each of these groups.  These counts constitute the elements of the vector $n$.  The $n$ vector is paired with a hypothetical $c$ vector, and an example property is calculated.

One of the earliest examples of GC methods is from Benson and Buss~\cite{bb58}, who are considered to be the originators of so-called ``group increment theory.''  Group increment theory, or Benson group increment theory (BGIT), is analogous to GC methods, but these terms may be more common in the physical chemistry literature.  In the original 1958 paper~\cite{bb58}, Benson and Buss proposed a simple group additivity scheme for the prediction of bond dissociation energies. Benson et al.~\cite{bcg+69} extended this work to account for a greater diversity of groups and to estimate heat capacities.  Additional work from Cohen and Benson includes estimating heats of formation with group increment theory~\cite{cb93}.  A large number of additional efforts have used Benson-like increments to estimate the same thermophysical properties, a very small sample of which are~\cite{dh88,jd86,rpe+05}.

Another very popular GC method was devised by Joback and Reid~\cite{jr87}.  This method extended the group increment idea to model many different properties with the same set of groups.  The Joback and Reid model also included functional transformations for the original group increment summations.  These altered the group contribution definition to the following:
\begin{equation}
P=f\left(\sum\limits_g c_g n_g\right)
\end{equation}
where $f$ represents some function of the inner product of the vectors $c$ and $n$.  These functions $f$ appear in many group contribution methods and are important when predicted properties are not simple linear functions of the number of groups in a structure.

Perhaps the most widely-used GC method in CAMD is that of Marrero and Gani~\cite{mg01,mg02}.  This method, like the method of Constantinou and Gani~\cite{cg94} that predated it, provides another extension to the general form of GC methods in that it introduces multiple levels of groups to better capture proximity effects, meaning the effect of two or more groups which are close to one another in a molecular structure.  As many GC methods are limited to groups of just a few atoms and bonds, many cannot differentiate between structures with different connectivity.  The Marrero-Gani method, called the GC+ method, uses as a first-order approximation a normal group contribution method, where the groups belong to a set of primary groups $F$.  An additional set of groups $S$ contains slightly larger sub-structures.  Finally, a set of groups $T$ accounts for large groups and overarching molecular structural features.  Unlike groups in the set $F$, groups in the sets $S$ and $T$ are allowed to overlap with each other.  Using the hierarchical depiction of molecules with the GC+ method, a much clearer picture of a molecule is provided.  The general form of the GC+ estimates is shown below:
\begin{equation}
P=f\left( \sum\limits_{g \in F} c_g n_g + \sum\limits_{g \in S} c_g n_g +\sum\limits_{g \in T} c_g n_g  \right)
\end{equation}

Group contribution methods have also often incorporated interaction terms~\cite{nrrc04,nrr07,nrr08,nrr09,kr84,pabh00}.  These are ways to include additional terms to account for the simultaneous presence of two (same or different) groups in a particular structure.  For example, in predicting toxicity, one group in a molecule may lead to a simple metabolic pathway and therefore make many structures containing that group non-toxic.  If that structure were also to have another group which is normally quite toxic, a GC method without interaction terms may not predict toxicity well.  This would be because both groups would have an additive effect, meaning that there would be one highly non-toxic contribution and one highly toxic contribution.  As a result, the molecule may be predicted to have an average toxicity when, in reality, the presence of the non-toxic group should outweigh the toxic group.  Introducing an interaction term for these two groups accounts for the situation of their co-occurrence.  In this example, this interaction term would likely remove whatever toxicity value was predicted by the toxic group.  For more discussion of interaction terms in predicting toxicity, readers are referred to ~\cite{my01}.  Interaction terms generally take the form
\begin{equation}
I_{g,g'}=f_I\left(n_g, n_{g'}\right)
\end{equation}
where $f_I$ usually represents multiplication, but can also represent other functions.  Group contribution methods can also include some idea of structural features~\cite{nrrc04,nrr07,nrr08,nrr09,mg01,mg02}.  These account for larger effects, typically at the molecule scale, such as aromatic ring substitution, cis/trans isomerism, aliphatic chain lengths, etc.  These can typically be implemented as large groups but sometimes require special considerations.  Table~\ref{gc_methods_table} provides several properties typically used in CAMD problems and a few GC methods to estimate them.

\textbf{\textit{Strengths of GC methods.}} GC methods are useful in that they are very intuitive to use.  They represent a chemical structure in terms of its functional components, very analogously to how chemists compare and analyze structures.  GC methods are also able to easily represent a large and diverse chemical space as the groups can be combined in many different ways to produce a large variety of different structures.  This is especially useful from a CAMD perspective.  Finally, GC methods are easily translated into the mathematical formulations of CAMD problems as the inclusion and count of the groups (the vector $n$) are easily represented in the context of mathematical optimization.

\textbf{\textit{Weaknesses of GC methods.}} There are a few shortcomings of modern GC methods.  One is that many GC methods are unable to distinguish isomers from one another.  As isomers can have very different properties, this represents a gap in the predictive power of GC methods.  We note that some GC methods such as the GC+ methods are able to distinguish many isomers due to the inclusion of large groups.  A second issue with GC methods is the lack of consistency in groups used to predict various properties.  Though this has no major effect estimating these properties for a given structure, it becomes problematic for mathematical formulations of the CAMD problem.  Finally, GC methods require specifying the set of groups prior to regressing the GC coefficients.  Though many GC methods are quite accurate, there is no guarantee that the set of groups used best captures the property they model.  Using different groups can sometimes drastically alter the predictive power of a GC model.

\begin{table*}[ht]
\footnotesize
\caption{Sample of available GC methods for predicting various properties of pure compounds\label{gc_methods_table}}
\begin{tabular}{ L{5cm} L{10cm} }
\textbf{Property} & \textbf{GC methods} \\
\hline
Aqueous solubility & Marrero et al.~\cite{mg02}, Klopman~\cite{kz01} \\
\hline
Boiling point &  Joback and Reid~\cite{jr87}, Stein et al.~\cite{sb94}, Nannoolal et al.~\cite{nrrc04}, Marrero et al.~\cite{mg01}  \\
\hline
Bond dissociation energy & Benson et al.~\cite{bb58} \\
\hline
Critical pressure    &  Jalowka et al.~\cite{jd86}, Joback and Reid~\cite{jr87}, Klincewicz et al.~\cite{kr84}, Nannoolal et al.~\cite{nrr07}, Marrero et al.~\cite{mg01}     \\
\hline
Critical temperature &  Jalowka et al.~\cite{jd86}, Joback and Reid~\cite{jr87}, Klincewicz et al.~\cite{kr84}, Nannoolal et al.~\cite{nrr07}, Marrero et al.~\cite{mg01}      \\
\hline
Critical volume &   Klincewicz et al.~\cite{kr84}, Nannoolal et al.~\cite{nrr07}, Marrero et al.~\cite{mg01}     \\
\hline
Enthalpy of formation   & Cohen et al.~\cite{cb93}, Benson~\cite{b99}, Domalski et al.~\cite{dh88}, Roganov et al.~\cite{rpe+05}, Joback and Reid~\cite{jr87}, Marrero et al.~\cite{mg01}         \\
\hline
Enthalpy of fusion & Joback and Reid~\cite{jr87}, Marrero et al.~\cite{mg01}   \\
\hline
Enthalpy of vaporization & Roganov et al.~\cite{rpe+05}, Joback and Reid~\cite{jr87}, Marrero et al.~\cite{mg01}, Ceriani et al.~\cite{cgm09}    \\
\hline
$\text{LC}_\text{50}$   & Martin et al.~\cite{my01} (fathead minnow) \\
\hline
Melting point &  Joback and Reid~\cite{jr87}, Marrero et al.~\cite{mg01}  \\
\hline
Gibbs energy of formation & Cohen et al.~\cite{cb93}, Benson~\cite{b99}, Domalski et al.~\cite{dh88}, Roganov et al.~\cite{rpe+05}, Joback and Reid~\cite{jr87}, Marrero et al.~\cite{mg01}         \\
\hline
Heat capacity		& Benson et al.~\cite{bcg+69}, Benson~\cite{b99}, Domalski et al.~\cite{dh88}, Joback and Reid~\cite{jr87},  Kolsk{\'a} et al.~\cite{kkzr08}, Ceriani et al.~\cite{cgm09}      \\
\hline
Octanol/water partition coefficient & Marrero et al.~\cite{mg02}, Platts et al.~\cite{pabh00}, Klopman et al.~\cite{klwd94} \\
\hline
Vapor pressure & Nannoolal et al.~\cite{nrr08} \\
\hline
Viscosity & Joback and Reid~\cite{jr87}, Sastri et al.~\cite{sr92}, Ceriani et al.~\cite{cgr+97}, Cao et al.~\cite{ckf+93}, Nannoolal et al.~\cite{nrr09} \\
\hline
\end{tabular}
\end{table*}

\subsection{Topological indices}

Chemical graph theory~\cite{chemical_graph_theory} is a field which became very influential in the 1970's and has since been used to produce a large number of QSPRs.  The basic idea of chemical graph theory is that the atoms and bonds which constitute a molecule can be thought of as nodes and edges in a graph.  In general, we use $G=(V,E)$ to define a graph $G$, with vertex set $V$ and edge set $E$.  Using this depiction of molecular structures, various properties of that graph, referred to here as topological indices, can be used as descriptors in QSPR models.  More specifically, various topological indices are paired with regression coefficients and used to estimate properties in a similar way to GC methods.

\begin{figure*}
\centering
\caption{Randi\'{c} edge indices for a simple alkane\label{ci_alkane}}
\footnotesize
\begin{tabular}{c c}
\textbf{Structure} & \textbf{Calculating $CI_E(v,v')$} \\
\hline
\begin{tabular}{c}
\includegraphics[scale=2.2]{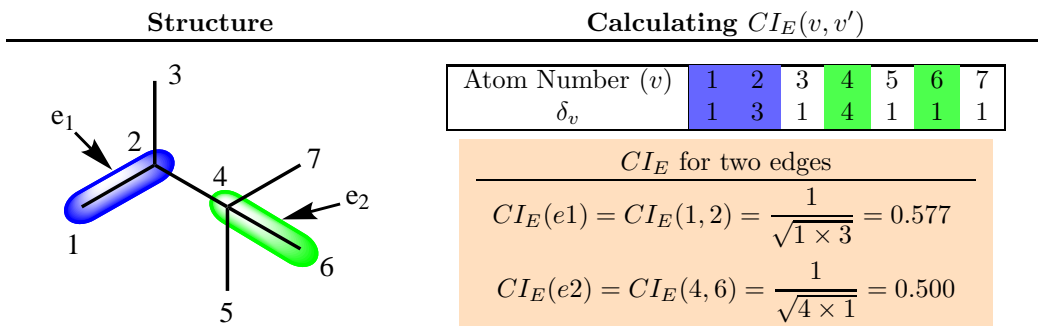}
\end{tabular}
&
\begin{tabular}{c}

\begin{tabular}{cccccccc}
\hline
\multicolumn{1}{|c|}{Atom Number ($v$)} &\cellcolor{blue!60!white}1&\cellcolor{blue!60!white}2&3&\cellcolor{green!70!white}4&5&\cellcolor{green!70!white}6&\multicolumn{1}{c|}{7} \\
\multicolumn{1}{|c|}{$\delta_v$} &\cellcolor{blue!60!white}1&\cellcolor{blue!60!white}3&1&\cellcolor{green!70!white}4&1&\cellcolor{green!70!white}1&\multicolumn{1}{c|}{1} \\ \hline  \vspace{-3mm} \\

\end{tabular}
\\
\colorbox{orange!25!white}{
\begin{tabular}{c}
$CI_E$ for two edges \\
\hline \noalign{\vskip 1mm}
$\displaystyle CI_E(e1)=CI_E(1,2)=\frac{1}{\sqrt{1 \times 3}}=0.577$ \vspace{2mm}\\
$\displaystyle CI_E(e2)=CI_E(4,6)=\frac{1}{\sqrt{4 \times 1}}=0.500$
\end{tabular}
}
\end{tabular}
\end{tabular}
\end{figure*}

Topological indices (TIs) can take many forms.  They are defined as some function of the nodes and edges in a chemical graph, and one can easily see that there are a large number of possible functions even just considering standard molecular graph properties like degree counts for nodes, connectivity, atomic types, etc.  One of the first topological indices used in chemical graph theory is the Wiener index~\cite{w47}.  The Wiener index attempts to describe the total distance between all atoms in the graph, as given by $d(v,v')$, the graph theoretic distance between all pairs of vertices $v$ and $v'$.  The Wiener index $W(G)$ is defined as:
\begin{equation}
W(G) = 1/2 \sum\limits_{v,v'} d(v,v')
\end{equation}
While the Wiener index describes a graph in terms of its distances, another important consideration is how a graph is connected.  To address this, an important class of topological indices called connectivity indices (CIs) was developed.  Connectivity indices are widely used in CAMD and have been shown to be useful in QSPR applications~\cite{er99}.  The first connectivity indices were developed by Randi\'{c}~\cite{r75}, who used these indices to account for the degree of branching in alkanes and to model enthalpy of fusion and vapor pressure.  Randi\'{c} defined an edge index to be:
\begin{equation}
CI_E(v,v') = \frac{1}{\sqrt{\delta_v \delta_{v'}}}
\end{equation}
where $v$ and $v'$ are two connected vertices in the chemical graph.  This means that the atoms which correspond to $v$ and $v'$ are connected by a chemical bond.  Furthermore, $\delta_v$ and  $\delta_{v'}$ are the degrees of nodes $v$ and $v'$.  In the study of Randi\'{c}, these degrees signified the number of $\sigma$ bonds a particular atom had to non-Hydrogen atoms (i.e., the number atomic neighbors in the hydrogen-suppressed graph), but they are sometimes defined differently for other connectivity indices.  The connectivity index of the entire molecule (graph) was then given by
\begin{equation}
^1 \chi = \sum\limits_{\{v,v'\} \in E}CI_E(v,v') = \sum\limits_{\{v,v'\} \in E} \frac{1}{\sqrt{\delta_v \delta_{v'}}}
\end{equation}
where $E$ is the edge set of the graph.  A calculation for the edge connectivity indices for a simple alkane is shown in Fig.~\ref{ci_alkane}.

$^1 \chi$ is the so-called first-order Randi\'{c} connectivity index.  An even simpler ``connectivity index'' exists which does not account for bonding at all.  This is called the zeroth-order connectivity index, and is given below:
\begin{equation}
^0 \chi = \sum\limits_{v \in V} \frac{1}{\sqrt{\delta_v }}
\end{equation}
Kier et al.~\cite{khmr75}, Hall et al.~\cite{hkm75}, and Murray et al.~\cite{mhk75} were the first to apply these connectivity indices as descriptors in QSPR models.  These models are typically linear in the descriptors, but many variations exist.  An important step forward came from Kier et al.~\cite{kmwrh76} who introduced higher-order connectivity indices.  In general, these are defined for an index of order $i$ as:
\begin{equation}
^i \chi = \sum\limits_{\{v_1,v_2,\hdots\} \in V_C^i } \frac{1}{\sqrt{\prod\limits_v \delta_v }}
\end{equation}
where $V_C^i $ is the set of all sets of $i$ connected vertices.  Kier and Hall~\cite{kh76} developed additional connectivity indices to account for heteroatoms.  These modified connectivity indices distinguished atoms by their valence electrons, leading to a new value for vertex degree, $\delta^\text{V}$, where the superscript indicates that valence is considered.  For second period elements, Kier and Hall~\cite{kh76} define $\delta_v^\text{V}$ for a vertex $v$ as
\begin{equation}
\delta_v^\text{V}= Z_v^\text{V} - h_v
\end{equation}
where $Z_v^\text{V}$ indicates the number of valence electrons for atom/vertex $v$ and $h_v$ is the number of hydrogens attached to $v$.  For atoms in the third period and beyond, the main difference from this perspective is the number of core electrons.  These are accounted for in the following:
\begin{equation}
\delta_v^\text{V}= \frac{Z_v^\text{V} - h_v}{Z_v-Z_v^\text{V}-1}
\end{equation}
where $Z_v$ is the atomic number of atom $v$.  These modified $\delta^\text{V}$ values define analogous connectivity indices $^i \chi^\text{V}$.

The connectivity index $\chi$ can be thought to primarily capture vertex adjacency and the local neighborhoods of every atom in a molecule.  As such, it represents how atoms are connected and perhaps not the ensemble molecular structure and shape.  Another topological index often used in CAMD that aims to address this is the so-called shape index, $\kappa$.  The shape index accounts for features of the entire molecular structure as functions of the underlying graph architecture and counts of graph substructures.  The main parameter in the calculation of $\kappa$ is the number of paths in the chemical graph of a certain length.  We define $^iP$ to be the number of paths of length $i$ in a particular chemical graph.

\begin{figure*}
\caption{Getting $^iP$ and calculating $^i\kappa$ for an example molecule\label{shape_index_example}}
\footnotesize
\centering
\begin{tabular}{c}
\hline
\begin{tabular}{C{3.4cm}C{3.4cm}C{3.4cm}C{3.4cm}}
Number of atoms & Paths of length 1 & Paths of length 2 & Paths of length 3 \\
\includegraphics[scale=1.5]{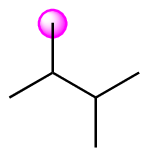} &
\includegraphics[scale=1.5]{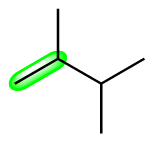} &
\includegraphics[scale=1.5]{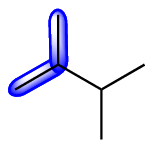} &
\includegraphics[scale=1.5]{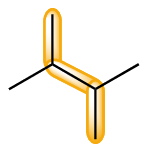} \\
$^0P$: 6 & $^1P$: 5 & $^2P$: 6& $^3P$: 4 \\
\end{tabular} \vspace{1mm}\\
\colorbox{orange!25!white}{
\begin{tabular}{C{3cm} c}
Calculating $\kappa$ values &
\begin{tabular}{ccc}
 $\displaystyle ^1\kappa =\frac{ 2 (^1P_\text{max}) (^1P_\text{min})}{  (^1P)^2 }$
& $\displaystyle ^2\kappa =\frac{ 2 (^2P_\text{max}) (^2P_\text{min})}{  (^2P)^2 }$
& $\displaystyle ^3\kappa =\frac{ 4 (^3P_\text{max}) (^3P_\text{min})}{  (^3P)^2 }$ \\
 $\displaystyle  =\frac{ 2 (30) (5)}{ 5^2 }=12$
& $\displaystyle  =\frac{ 2 (10) (4)}{  6^2 }=2.22$
& $\displaystyle =\frac{ 4 (4) (3)}{  4^2 }=3$ \\
\end{tabular}
\end{tabular}
}\\
\hline
\noalign{\vskip 0.3mm}
\begin{tabular}{cccccccccc}
\vspace{0.5mm}
Other structures &$^0P$ & $^1P$& $^2P$& $^3P$
&Other structures &$^0P$ & $^1P$& $^2P$& $^3P$ \\
\hline
\adjustbox{valign=c}{\includegraphics[scale=1.5]{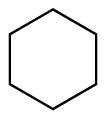} }& 6 & 6 &6 & 6
&\adjustbox{valign=c}{\includegraphics[scale=1.5]{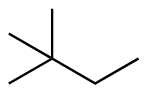} }& 6 & 5 &7 & 3 \\
\adjustbox{valign=c}{\includegraphics[scale=1.5]{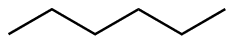} }& 6 & 5 &4 & 3
&\adjustbox{valign=c}{\includegraphics[scale=1.5]{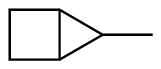} }& 6 & 7 &11 & 13 \\

\end{tabular}
\end{tabular}
\end{figure*}

Like other topological indices, the shape index maps this information onto a single value.  There are three often used $^i\kappa$ values which are defined as:
\begin{align}
\displaystyle ^1\kappa =\frac{ 2 (^1P_\text{max}) (^1P_\text{min})}{  (^1P)^2 } \\
^2\kappa =\frac{ 2 (^2P_\text{max}) (^2P_\text{min})}{  (^2P)^2 } \\
 ^3\kappa =\frac{ 4 (^3P_\text{max}) (^3P_\text{min})}{  (^3P)^2 }
 \end{align}
with $^iP_\text{max}$ and $^iP_\text{min}$ representing the maximum and minimum possible number of paths of length $i$ for a hypothetical molecule with an equivalent number of atoms. $^iP_\text{max}$ and $^iP_\text{min}$ can be easily derived from graph theoretic arguments.  The formal expressions for these can be found in~\cite{hk07}. Several $^i\kappa$ values for an example molecule are calculated in Fig.~\ref{shape_index_example}.

In Table~\ref{ti_table}, we list a few references for QSPR models using topological indices.  There are many, many examples of such models, and we note that this table is not meant to be exhaustive.  For a more complete list of available topological-indices-based QSPRs, the reader is directed to the book of Devillers and Balaban~\cite{top_index_book}.

\begin{table*}[ht]
\footnotesize
\caption{Sample of available TI-based methods for predicting various properties of pure compounds\label{ti_table}}
\begin{tabular}{ L{5cm} L{10cm} }
\textbf{Property} & \textbf{TI method} \\
\hline
Anti-inflammatory activity & Gupta et al.~\cite{gsm02}, Bajaj~\cite{bsm05} \\
\hline
Aqueous diffusion coefficient & Schramke~\cite{smdwh99} \\
\hline
Aqueous solubility & Hall et al.~\cite{hkm75}, Katritzky et al.~\cite{kwstk98}  \\
\hline
Biodegradability & Boethling~\cite{b86} \\
\hline
Blood-brain barrier partition coefficients & Rose et al.~\cite{rhk02} \\
\hline
Boiling point & Hall et al.~\cite{hkm75}, Hosoya~\cite{h71}, Hall et al.~\cite{hs96}, Galvez et al.~\cite{ggss94} \\
\hline
Critical temperature & Hall et al.~\cite{hs96} \\
\hline
Density & Kier et al.~\cite{kmwrh76}, Estrada~\cite{e95}, Katritzky et al.~\cite{kg93} \\
\hline
Enthalpy of formation   &  Mercader~\cite{mct00}    \\
\hline
Enthalpy of fusion &  Gharagheizi et al.~\cite{ggv12}  \\
\hline
Enthalpy of vaporization & Galvez et al.~\cite{ggss94}  \\
\hline
Flash point & Patel et al.~\cite{pnm09} \\
\hline
Heat capacity		&  Yao et al.~\cite{yfdplzzh03}     \\
\hline
$\text{LC}_\text{50}$(fathead minnow) & Hall et al.~\cite{hmk89,hmk89_other}, Basak et al.~\cite{bgm84} \\
\hline
Melting point & Katritzky et al.~\cite{kg93}   \\
\hline
Nonspecific local anesthetic activity & Kier et al.~\cite{khmr75}, Katritzky et al.~\cite{kg93} \\
\hline
Octanol/water partition coefficient & Murray et al.~\cite{mhk75} \\
\hline
$\pi$-electron energy (C-C bonds) & Hosoya et al.~\cite{hhg75} \\
\hline
Refractive index & Katritzky et al.~\cite{kg93} \\
\hline
Vapor pressure & Katritzky et al.~\cite{kwstk98} \\
\hline
Viscosity &  Kauffman et al.~\cite{kj01} \\
\hline
Water-air partition coefficient &Katritzky et al.~\cite{kwstk98}  \\
\hline
\end{tabular}
\end{table*}

Table~\ref{ti_table} provides some idea of the diversity of applications of QSPR development with topological indices.  We note that this table is inclined towards properties common in CAMD problems, so the properties listed may be more relevant to chemical engineering applications.  However, we provide a few examples of the many applications of topological indices to modeling biological, environmental, and pharamacological properties.  There have been many efforts to build models related to pharmaceutical properties~\cite{basak87,eu01}, so only a small subset is listed here.  One additional interesting application of connectivity indices is in the prediction of coefficients for group contribution methods where the group is missing (the group is not in the descriptor space)~\cite{ghh05}.

\textbf{\textit{Strengths of TIs.}} One of the main advantages of TIs is that they can discriminate between very similar structures, often in cases where GC methods cannot (e.g., isomers).  This provides a more holistic picture of the molecule and can be very useful for certain design problems.  For example, this may have potential applications for CAMD in areas where a structural feature of the to-be-designed compound is fixed \textit{a priori}. Furthermore, since many TIs are a function of the entire graph, TIs reflect the entire nature of the molecular structure.  This can have advantages over GC methods, which assume that each group provides a contribution independently of other groups in the structure (this is offset somewhat by GC interaction functions).  Finally, TIs have been extensively applied to modeling pharmacological properties.  The quality and volume of this literature means that TIs are very suitable to many pharmaceutically relevant CAMD problems.

\textbf{\textit{Weaknesses of TIs.}}  Though topological indices have been widely applied to QSPR, there are only limited examples in CAMD.  Topological indices are usually not as generally applicable as GC methods, meaning that TI-based QSPRs are often restricted to a certain class of chemicals.  For design purposes, this is problematic as it means that TI-based design problems can only consider that particular subset of the chemical search space.  Furthermore, TIs represent graph-theoretic properties of the chemical graph, and many of these properties are not always readily understandable from a chemical perspective (although Randi\'{c} and Zupan~\cite{rz01} have offered some interpretations of several TIs).  Finally, TIs are more difficult to incorporate into CAMD than GC methods.  We will discuss TI-based CAMD in a subsequent section, but such CAMD problems can sometimes face combinatorial difficulties and have only been demonstrated thus far on small design problems.

\subsection{Signature descriptors}

In a broad sense, GC methods capture the important subsets of atoms in a molecule while TIs rely on some function of the chemical graph.  One QSPR method which has been shown to capture aspects of both GC- and TI-based methods is signature descriptors (SD).  Signature descriptors are far younger than the other methods discussed, originating in 2002 from Visco et al.~\cite{vprf02} and 2003 from Faulon et al.~\cite{fvp03}. Like TIs, they conceive of chemical structures as the chemical graph.  Rather than ascribe various values to a complete molecular graph, SDs retain all of the structural and connectivity information for every atom in a molecule.

Analogously to TIs, SDs define the chemical graph to be $G=(V,E)$.  Standard SD methods also incorporate node coloring for every node $v$ via a coloring function $c_G(v)$ and the colors of each node $C_v$.  This change is reflected in the slightly altered definition of the chemical graph as $G=(V,E,C,c_G)$.  These node colorings are intended to distinguish between different atoms as well as different types of the same atom.  For example, it may be beneficial for the model to differentiate between an oxygen with two single bonds and an oxygen with a double bond.  Additionally, one may want to distinguish hydrogens by what atom they are attached to, aromatic carbons from non-aromatic ones, atoms attached to aromatic rings, and many other chemical features.  We note that the colorings of the nodes are one of the only subjective parts of SD models, and different coloring schemes can have significant effects on the performance of the models.

\begin{figure*}
\caption{Atomic signature descriptors for a carbon atom in an example molecule\label{sd_example}}
\footnotesize
\centering
\begin{tabular}{C{3.5cm} c C{4.5cm}}
\textbf{Molecular representation} & \textbf{Tree representation} & \textbf{Atomic signature descriptors}\\
\hline
\adjustbox{valign=c}{\includegraphics[scale=1]{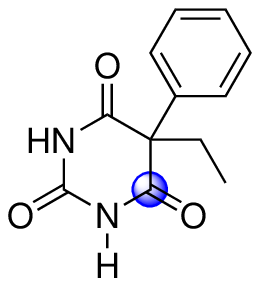}}
&
\begin{tikzpicture}[baseline=(current bounding box.center),
    level0/.style={circle, draw=none, rounded corners=1.5mm, fill=blue!70!white,
        text centered, anchor=north, text=white},
    level1/.style={circle, draw=none, fill=red,
        text centered, anchor=north, text=white},
    level2/.style={circle, draw=none, fill=green,
        text centered, anchor=north, text=white},
    level distance=0.8cm, growth parent anchor=south
]
\node (Lv0) [level0,label={[align=center]below:C\\$sp^2$}] {} [-];
\end{tikzpicture}
&
\colorbox{blue!40!white}{
\begin{tabular}{c}
\hline
Height 0 descriptor \\
\hline
\noalign{\vskip 0.3mm}
$\text{C}^2$
\end{tabular}
}
\\
\adjustbox{valign=c}{\includegraphics[scale=1]{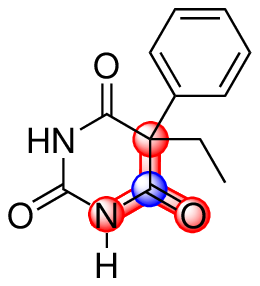}}
&
\adjustbox{valign=c}{
\begin{tikzpicture}[baseline=(current bounding box.center),
    level0/.style={circle, draw=none, rounded corners=1.5mm, fill=blue!70!white,
        text centered, anchor=north, text=white},
    level1/.style={circle, draw=none, fill=red!90!white,
        text centered, anchor=north, text=white},
    level2/.style={circle, draw=none, fill=green,
        text centered, anchor=north, text=white},
    level distance=0.3cm, growth parent anchor=south
]
\node (Lv0) [level0] {    } [-]
    child{
        node (lv1_1) [level1,label={[align=center]below:N\\$sp^3$} ] {}
        }
     child{
     	node (lv1_2) [level1,label={[align=center]below:O\\$sp^2$}] {}}
     	child{
     	node (lv1_3) [level1,label={[align=center]below:C\\$sp^3$}] {}
            };
\end{tikzpicture}
}
&
\colorbox{red!40!white}{
\begin{tabular}{c}
\hline
Height 1 descriptor \\
\hline
\noalign{\vskip 0.3mm}
$\text{C}^2(\text{N}^3)(\text{O}^2)(\text{C}^3)$
\end{tabular}
}
\\
\adjustbox{valign=c}{\includegraphics[scale=1.0]{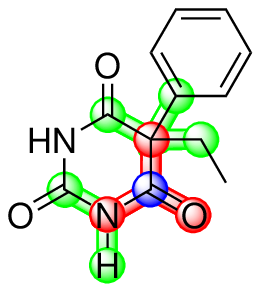}}
&
\begin{tikzpicture}[baseline=(current bounding box.center),
    level0/.style={circle, draw=none, rounded corners=1mm, fill=blue!70!white,
        text centered, anchor=north, text=white},
    level1/.style={circle, draw=none, fill=red!90!white,
        text centered, anchor=north, text=white},
    level2/.style={circle, draw=none, fill=green!90!white,
        text centered, anchor=north, text=white},
    level2a/.style={circle, draw=none, fill=green!90!white,
        text centered, anchor=east, text=white},
    level distance=0.3cm, growth parent anchor=south
]
\node (Lv0) [level0] {    } [-]
    child{
        node (lv1_1) [level1] {}
        child{
            node (lvl2_1) [xshift=0.3cm,level2,label={[align=center]below:C\\$sp^2$}] {}
            }
        child{
            node (lvl2_1) [xshift=-0.2cm,level2,label={[align=center]below:$\text{H}_\text{A}$}] {}
            }}
        child{
     	      node (lv1_2) [level1] {}}
     	child{
     	      node (lv1_3) [level1] {}
     	      child{
                   node (lvl2_2) [xshift=-0.0cm,level2,label={[align=center]below:C\\$sp^2$}] {}
                      }
              child{
                   node (lvl2_3) [xshift=-0.4cm,level2,label={[align=center]below:aC}] {}
                      }
              child{
                   node (lvl2_2) [xshift=-1.05cm,level2,label={[align=center]below:C\\$sp^3$}] {}
                      }
            };
\end{tikzpicture}
&
\colorbox{green!40!white}{
\footnotesize
\begin{tabular}{c}
\hline
Height 2 descriptor \\
\hline
\noalign{\vskip 0.3mm}
\begin{tabular}{c}
$\text{C}^2(\text{N}^3(\text{C}^2)(\text{H}_\text{A}))$ \\
$(\text{O}^2)$\\$(\text{C}^3(\text{C}^2)(\text{aC})(\text{C}^3))$
\end{tabular}
\end{tabular}
}
\end{tabular}
\end{figure*}

One important class of signature descriptors is known as atomic signatures.  Given a certain atom in a chemical graph, its atomic signature represents all of the atoms that are within a certain distance, or height, from it.  Varying the value of this distance gives rise to different atomic signatures.  In the simplest case, with this distance set to 0, the atomic signature of an atom is simply that atom, colored in keeping with the coloring definition.  More formally, for an atom (vertex) $v$, its atomic signature of height 0, $^0\alpha$ is given by:
\begin{align}
^0\alpha(v) = c_G(v)
\end{align}
Of course, these atomic signatures of height 0 are not a robust set of descriptors.  Considering higher values for height provides a more detailed picture of an atom's environment.  In general, for a height of $i$, the atomic signature of height $i \geq 1$ for a vertex $v$ is defined as:
\begin{align}
^i\alpha(v) = G_S(V^i,E^i,C^i,c_G)
\end{align}
where $G_S$ defines a subgraph of $G$ that contains all vertices and bonds such that the distance between $v$ and any vertex in $G_S$ is at most $i$.  Thus, an atomic signature of height one for a hypothetical atom $v$ defines $v$ and every atom bonded to $v$ as well as all connecting edges.  A height two atomic signature defines $v$, every atom bonded to $v$, and every atom bonded to those atoms bonded to $v$ along with all necessary edges.  Increasing the height of an atomic signature can thus be thought to add a layer of connected atoms.  In graph-theoretic terms, increasing the $i$ value is equivalent to adding a layer to a breadth-first-search.  A pictorial explanation is provided in Fig.~\ref{sd_example}.  In this example, we assume that all atoms are colored by aromaticity and $sp$ hybridization and that hydrogens are colored by the atom they are attached to.

The atomic signatures can be thought of as a descriptor space for a molecular structure.  A property, $P$, of a molecule can be estimated by all of its atomic signatures of up to a particular height.  This QSPR has a familiar form:
\begin{align}
P= \sum\limits_i \sum\limits_{d \in D^i} c_d~^i\alpha_G(d)
\end{align}
where $d$ is the index of the set of all atomic signatures and $D^i$ is the set of atomic SDs of height $i$.  $c_d$ is a regression coefficient accounting for the ``contribution'' of each atomic signature to a certain property.  $^i\alpha_G(d)$ represents the number of occurrences of atomic signature $d$.  Using the atomic signatures, signatures of the entire molecule can also be generated.

A major advantage of signature descriptors is that they can be manipulated via simple functions to represent groups from GC methods as well as various TIs.  This means that the large amount of QSPRs derived from GC methods and TI-based methods are accessible using SDs.  For this reason, we do not provide a table of QSPRs using SDs because they can be---and often are---used to calculate properties via GC- and TI-based methods.  A few examples of converting SDs to groups and TIs are given in~\cite{fvp03}.

\textbf{\textit{Strengths of SDs.}} One of the main advantages of signature descriptors is that they have a small inherent bias as compared to GC methods and TIs.  The only bias introduced in these descriptors is the choice of an atomic coloring scheme, or the choice of what defines a different type of the same atom.  Furthermore, the inclusion of every atomic signature in SD-based QSPRs means there are no theoretical restrictions on the descriptor space defined for a molecular structure.  There are also a variety of more modern signature descriptors capable of distinguishing stereoisomers.  In the case of stereoisomers, certain SDs have far greater discriminative power than GC methods or TIs, which typically cannot differentiate stereoisomers.  Furthermore, the equivalence of SDs to many TIs and groups for GC methods means that SDs can be used directly with these TI and GC QSPR models.  This makes a large library of QSPR models accessible to SDs.

\textbf{\textit{Weaknesses of SDs.}}  Many QSPRs using SDs use all of the available atomic signatures of up to a certain height.  This can quickly become an issue with the predictive power of SDs as models without sufficient training data may be overfit.  A second concern with SDs is that the coloring scheme of the descriptors must be specified before the descriptors are used for QSPRs and CAMD problems.  It is likely that some coloring schemes provide better results than others, and the best coloring scheme may not always be easy to determine.  Finally, atomic signature descriptors always discriminate between identical atoms in different structural environments.  This may again lead to issues of overfitting or not capturing the ``true'' chemical behavior of the system as sometimes it is better to model identical atoms with a general descriptor which is independent of the atom's environment (although this can be captured to some degree with SDs of lower heights).

\section{CAMD as an optimization problem}
\label{camd_optimization_prob}
The various types of structure-property relationships discussed above can quickly and often accurately estimate properties from a structure.  The application of QSPR techniques in this direction---\textit{predicting properties from structures}---defines what is known as the ``forward problem,'' and is what QSPR techniques are generally intended for.  CAMD can broadly be thought to consider the ``reverse problem,'' or the problem of \textit{predicting structures from properties}.  At first glance, there is no immediately obvious way of relating properties to a specific molecular structure.  One issue is that there are so many structures to consider.  A reasonable approach to the CAMD problem should be able to consider a large diversity of structures without running into significant computational difficulties.  Another issue involves structural feasibility.  Solutions to the CAMD problem must also be sensible molecular structures, meaning CAMD should produce structures that do not violate any inherent laws of chemical bonding.  A final issue involves consistency.  All of the QSPRs discussed (though non-overlapping GC methods are an exception) require that if certain features are present, so must be other features.  For example, it would be unreasonable to design a structure that has four paths of length three but no paths of length two.  In the case of GC methods or SDs, an example erroneous solution would contain one occurrence of the carboxylic acid ($\text{-C(=O)OH}$) group/signature and no occurrences of the carbonyl ($\text{-C(=O)-}$) group/signature.  Note that this example assumes that overlap is allowed between these groups/signatures (although this is generally the case with signatures).

Mathematical optimization is the key to addressing all of these issues.  Before introducing the problem in a general optimization formulation, we define a few important sets and variables.  First, we assume that we have a vector of properties $p$ and a property value $p_k$ for each property $k$.  The vector $n$ encapsulates relevant structural information of the designed molecules and is dependent on the type of QSPR chosen.  In the case of GC methods, the value $n_d$ would represent the number of occurrences of each group $d$.  For TIs, the $n$ vector may represent the number of topological features $d$ (edges, paths of certain lengths, etc.) from which TIs would be calculated.  For SDs, this $n$ vector usually represents counts of various atomic signature descriptors $d$.  The function $f$ then transforms this structural information into a property estimate using the appropriate QSPR relationship.  In general, the CAMD problem can then be expressed as:
\begin{align}
\underset{n}{\min} \quad & C(n,p) \label{camd_obj}\\
s.t. \quad  & p=f(n) \label{camd_qspr}\\
& h_1(p,n) \leq 0 \label{camd_ineq} \\
& h_2(p,n) = 0 \label{camd_eq} \\
& s_1(n) \leq 0 \label{camd_prop_ineq} \\
& s_2(n) = 0 \label{camd_prop_eq} \\
& p_k^L \leq p_k \leq p_k^U \quad \forall k\label{camd_p_bounds}\\
& n_d^L \leq n_d \leq n_d^U \quad \forall d \label{camd_n_bounds}
\end{align}
In the above, \eqref{camd_qspr} involves QSPR functions $f$ which estimate a vector of properties $p$ from attributes such as group counts, graph topological features, or atomic signatures.  \eqref{camd_ineq} \eqref{camd_eq} define general functions, $h_1$ and $h_2$, representing inequality and equality constraints on property values, desired structural features, process conditions, and a variety of other possibilities.  For example, the presence of certain groups or structural features may necessitate changing a chemical process to accommodate these structures.  These constraints can also eliminate certain groups/topological features/signatures from the solution space or require that they appear a certain number of times.  The functions $h_1$ and $h_2$ can also account for design considerations such as system thermodynamics, cost, and a variety of system-specific interactions between $n$ and $p$.  \eqref{camd_prop_ineq} and~\eqref{camd_prop_eq} define inequality and equality constraints which ensure structural feasibility.  More specifically, the functions $s_1$ and $s_2$ determine if the vector $n$ is consistent with a molecular structure which can actually exist.  These constraints prevent erroneous structures from being considered, eliminating compounds that violate atomic valences, are disjoint, have an incorrect number of aromatic atoms, etc.  \ref{camd_p_bounds} and~\eqref{camd_n_bounds} set bounds on property values and $n$, respectively.  Each property $k$ is bounded below by $p_k^L $ and above by $p_k^U$.  Similarly, $n_k^L $ and $n_k^U$ define lower and upper bounds for $n_k$.  Finally, \eqref{camd_obj} is a general objective function for the CAMD problem.  $C(n,p)$ can define a number of possible functions.  These functions somehow quantify the performance of a specific molecule based on its properties $p$ and perhaps its descriptors $n$.  Of course, this $C$ function can either be minimized or maximized depending on the design problem.

\subsection{Classes of the CAMD problem}

\subsubsection{Single molecule design}

The single molecule design problem is the problem of determining a single, optimal structure for a particular purpose.  For these problems, the structure of the compound is the only design consideration, meaning that the variables represented by $n$ above are the only degrees of freedom in the problem.  Furthermore, it is assumed that there exists some ranking criterion with which to determine which structures are better than others.  To make this section as general as possible, we assume that the ranking criterion can either be applied during the optimization procedure as the objective function or afterwards, evaluating the performance of each of a pool of candidate molecules.  Though many classifications of these single molecule design problems are possible, we suggest three basic forms: (1) determining all feasible structures; (2) using an objective function which directly quantifies a molecule's performance; and (3) designing a structure with properties as close as possible to certain property targets.

\begin{figure*}[t]
\caption{Pictorial representation of the problem of finding all feasible molecules\label{camd_feas_prob}}
\centering
\includegraphics[scale=1.0]{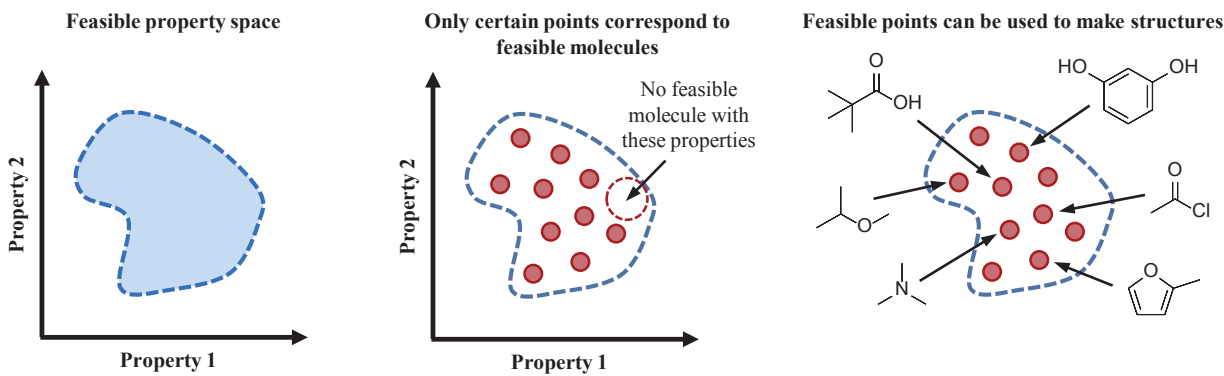}
\end{figure*}

\textit{The feasibility problem.}  We begin by describing the case where there is no objective function used in the optimization problem.  More formally, this is equivalent to solving the above CAMD formulation with the objective function equal to a constant.  The solutions to this problem represent all molecules which satisfy the functions $h_1$ and $h_2$, are chemically sound structures, and do not violate the property or descriptor constraints.  This is a useful type of problem to solve in CAMD when the performance function $C$ is not accessible or reliable.  It may be the case that $C$ requires a complex simulation or experimental work.  The performance function may also be inaccurate or the designer may not know exactly which properties to optimize.  In these cases, these CAMD problems leverage the power of optimization to reduce the large number of possible compounds to a more manageable number.  This smaller pool of compounds can then be investigated using high-order models or experiments.

Feasibility problems are sometimes difficult to distinguish from other types of problems.  In many cases, a feasibility problem is solved at one of the beginning stages of the problem, and all of the feasible structures are then evaluated based on some ranking criteria.  For these types of problems, there is no clear line between a feasibility problem and a problem with an objective function.  Usually, the molecules designed by these problems are ultimately ranked, so many ``feasibility problems'' actually have an implicit objective function.  A second complication is that many CAMD methodologies are designed to solve both feasibility problems and problems with objectives.  We provide a few examples of feasibility problems here for GC methods~\cite{joback89}, TIs~\cite{kh93,khd93}, and SDs~\cite{crmvklsbf04}.  We leave a more detailed discussion for the applications section of this article.  A graphical example of the feasibility problem is given in Fig.~\ref{camd_feas_prob}.

\begin{figure*}[t]
\caption{Pictorial representation of the problem of finding an optimal molecule\label{camd_opt_prob}}
\centering
\includegraphics[scale=1.0]{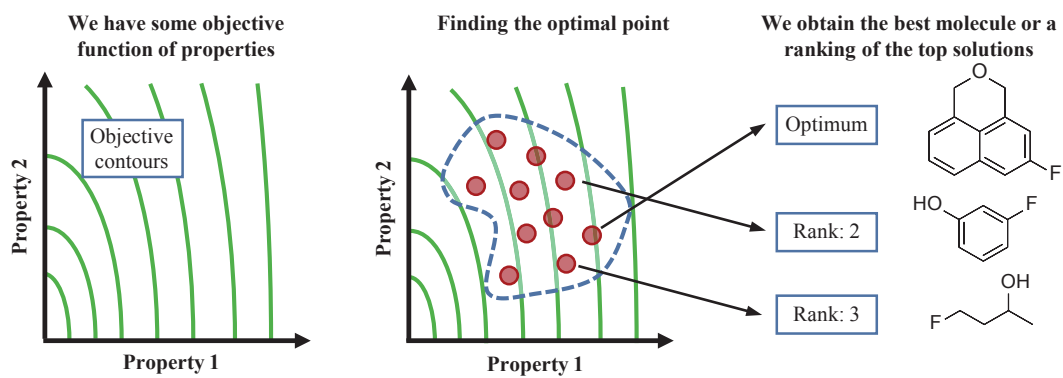}
\end{figure*}

\textit{Exact relationship between structures and performance.}  In many cases, it is possible to model a molecule's performance directly as a function of its structure and properties.  This is the most natural and common form of the CAMD objective because the performance function used to rank molecular structures is what is minimized or maximized.  Using the exact performance function as the objective $C(n,p)$ guarantees that the solution to the CAMD problem is the optimal molecule for the application, at least as judged by the provided performance function.  We also note that if the function $C(n,p)$ or any of the constraints is non-convex, the optimal molecule may only represent a local optimum.  Using a global optimization algorithm will provide the best molecule for these non-convex problems.

An exact representation of the objective has inspired many numerical optimization strategies to solving the CAMD problem.  Numerical optimization strategies are particularly advantageous in these cases because there is an exact algebraic relationship between the descriptor space $n$ and the performance function $C$.  This enables the use of modern, state-of-the-art optimization techniques.  Several of these will be discussed in an upcoming section.

There are many examples in the CAMD literature which fall into this category.  Again, we will postpone the discussion of these topics until later sections of this document.  We provide a small (and non-exhaustive) selection of relevant references in the meanwhile to offer some insight to the interested reader~\cite{om93,sao99,sty03}(GC), \cite{cm99,scts00} (TIs), and~\cite{csbe10} (SDs).  A graphical example of this type of CAMD problem is provided in Fig.~\ref{camd_opt_prob}.

\textit{Minimizing distance to property targets.}  Another class of single molecule design problems concerns finding a molecule with properties as close as possible to target values.  These types of formulations have been used extensively in CAMD, often to design alternatives to molecules being used in practice.  These molecules may need to be substituted for reasons of environmental friendliness, cost, or availability.  Many other types of problems can be addressed when the ideal properties of a molecule are known.  However, if these ideal properties do not represent a structure that is physically realistic, these approaches face some difficulty.

In general, these types of problems define the objective to be:
\begin{align}
C(p)=\sum\limits_k w_k (p_k^T - p_k)^2
\end{align}
where $p_k^T$ represents a targets for property $k$ and $w_k$ is a weight for the distance from from the estimated property value to its corresponding property target.  The second norm shown above is the most common distance function in CAMD, but many other distance functions are possible.

\begin{figure}
\caption{Pictorial representation of the problem of finding a feasible molecule with predicted properties as close as possible to target values \label{camd_target_prob}}
\centering
\includegraphics[scale=1.0]{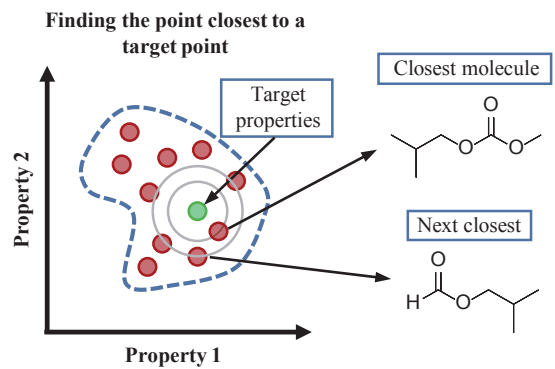}
\end{figure}

We present a few demonstrative examples of CAMD with property target objective functions.  Matsuda et al.~\cite{mykt07} used groups to design ionic liquids based on conductivity and viscosity targets.  Siddhaye et al.~\cite{scst04} addressed this problem with topological indices to design pharmaceutical products.  Brown et al.~\cite{bmrf06} used signature descriptors to design polymers with specific properties.  A graphical example of this type of CAMD problem is given in Fig.~\ref{camd_target_prob}.

\subsubsection{Mixture design}

Real-world applications often demand a product with specifically tailored properties.  This sometimes necessitates utilizing a mixture of compounds as no single compound possesses all of the necessary properties.  Using CAMD techniques to simultaneously design two or more compounds for use in a blend/mixture is referred to as the mixture design problem.  We note here that we alter a definition given in~\cite{asst16} for consistency with the prevailing definition of mixture design in the literature.  All applications classified under mixture design can be assumed to design two or more structures simultaneously.

\begin{figure*}[t]
\caption{Pictorial representation of the mixture design problem\label{mix_design_problem_graphic}}
\centering
\includegraphics[scale=1.0]{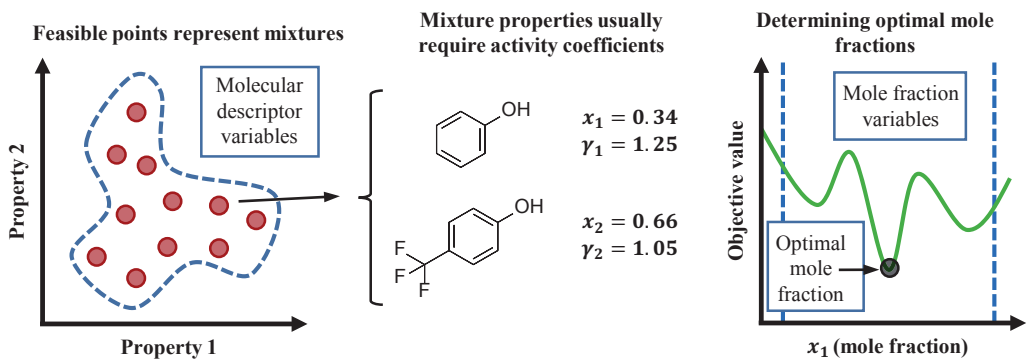}
\end{figure*}

While the single-molecule design is difficult in many cases, the mixture design problem is much harder.  The difficulties come from several sources: (1) the descriptor variables must now represent the descriptors of every unknown component in the mixture; (2) mixture properties must be calculated and included in the problem; (3) non-ideal mixture behavior must be considered in the form of thermodynamic relationships, activity coefficient models, or equations of state; (4) the design of mixtures also requires a determination of the amount of each component, so mole fractions must be considered.  We provide a pictorial representation of the mixture design problem in Fig.~\ref{mix_design_problem_graphic}.  In this problem, we define the variable $x_i$ to represent the mole fraction of unknown component $i$.  Furthermore, $q_j$ will represent the mixture property $j$.  The formulation of the mixture design problem is similar to the single-molecule design problem.  The objective function is altered to now include mixture properties:
\begin{align}
\underset{n,x}{\min} \quad C(n,p,q) \label{camxd_obj}
\end{align}
A few additional constraints are also necessary:
\begin{align}
q=g(x,n,p) \label{camxd_g} \\
\sum\limits_i x_i =1 \label{camxd_mole_fractions} \\
h_1(p,q,n) \leq 0 \label{camxd_ineq} \\
h_2(p,q,n) = 0 \label{camxd_eq} \\
q_j^L \leq q_j \leq q_j^U \label{camxd_q_bounds}
\end{align}
In the above, \eqref{camxd_ineq} and~\eqref{camxd_eq} represent modified constraints from the previous formulation to include the mixture property variables $q$.  Eq.\eqref{camxd_g} represents simple mixing functions or complex thermodynamic relationships which relate individual component property variables $p$, descriptor variables $n$, and mole fractions $x$ to the mixture property variables $q$.  In many cases, some of these $q$ values represent activity coefficients.  Eq.~\eqref{camxd_mole_fractions} ensures that mole fractions sum to 1.  Finally, \eqref{camxd_q_bounds} places upper ($q_j^U$) and lower ($q_j^L$) bounds on mixture properties.  We provide a few mixture design citations~\cite{gf93,cgn11,blp99} and leave the rest for the applications section.

\subsubsection{Integrated process and product design}

Though many CAMD endeavors design products with the ultimate goal of being incorporated into an industrial process, few have explicitly considered the relationship between a particular structure and a process.  This is especially important as process performance is typically very sensitive to the molecule(s) chosen.  These problems are especially challenging from an optimization point of view because there is no easily discernible algebraic relationship between the descriptor variables $n$ and process variables.  This problem also requires the introduction of process variables $\mu_w$, with $w$ defining the index over the process variables.  To present the formulation for integrated process and product design, we modify the single-molecule design problem for simplicity.

\begin{figure*}
\caption{A decomposition approach to the integrated product/process design problem\label{campd_graphic}}
\centering
\includegraphics[scale=1.0]{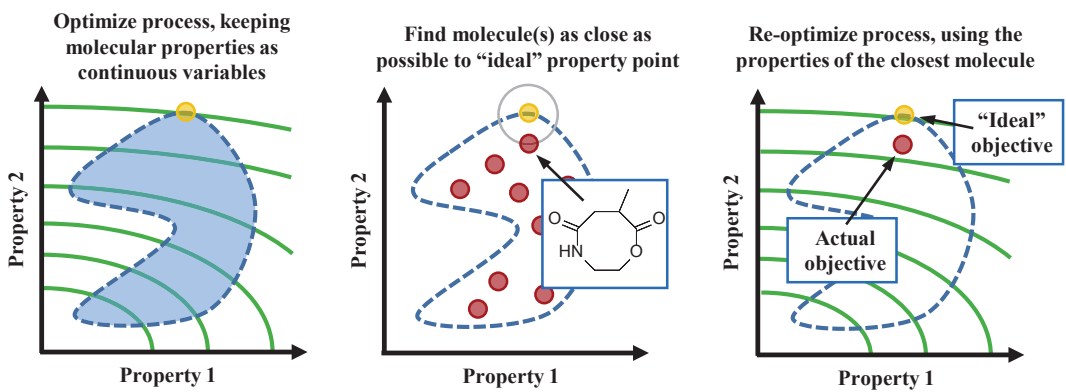}
\end{figure*}

First, the objective function now reflects process variables:
\begin{align}
\underset{n,\mu}{\min} \quad C(n,p,\mu) \label{campd_obj}
\end{align}
A few additional constraints then account for the inclusion of process variables:
\begin{align}
h_1(p,\mu,n) \leq 0 \label{campd_ineq} \\
h_2(p,\mu,n) = 0 \label{campd_eq} \\
\mu_w^L \leq \mu_w \leq \mu_w^U \label{campd_mu_bounds}
\end{align}
In the above, \eqref{campd_ineq} and~\eqref{campd_eq} represent modified constraints from the previous formulation to include the process variables $\mu$.  \eqref{campd_mu_bounds} is introduced to place upper ($\mu_w^U$) and lower ($\mu_w^L$) bounds on process variables.  There is a similar formulation defined for process/product design problems which also consider mixtures.  Again, we provide a small selection of references~\cite{ejge04,pl05,kag05} and leave the remaining discussion for the applications section.  One approach to the process design problem is illustrated in Fig.~\ref{campd_graphic}.

\subsection{Common design features in the form of constraints}

These constraints define any process criteria, design necessities, thermodynamic conditions, and any other features which may be important for a particular design problem.  As these constraints reflect the diversity of applications of CAMD, we cannot go into detail about all of them here.  Rather, we present a summary of design features and conditions which occur most commonly in CAMD.

A very common design feature in CAMD limits the number of groups which can occur in the solution.  For example, constraints of this variety can place a limit on the number of molecular descriptors $d$:
\begin{align}
n_d \leq n_d^U
\end{align}
or a limit on the total number of descriptors:
\begin{align}
\sum\limits_d n_d \leq N^{U}
\end{align}
where $N^U$ represents the maximum number of structural features (groups, atoms, descriptors, topological features, etc.) in the designed molecule.  Many design problems also set a lower bound on this summation, $N^L$, which is typically equal to 2.  This ensures that more than one descriptor appear in the solution.  Finally, some design problems seek solutions which are analogues within a given family of chemical structures.  In this case, it may be necessary to include the descriptor variables corresponding to this structural family:
\begin{align}
n_d^F \leq n_d \quad \forall d \in D^\text{FIX}
\end{align}
where $D^\text{FIX}$ defines the set of descriptors which must occur in the solution and $n_d^F$ represents the number of descriptors $d$ which must occur for the structural family to be produced.  This constraint is typically part of \eqref{camd_n_bounds}, where other descriptors may have lower bounds for other reasons.

Properties are also bounded similarly in equation~\eqref{camd_p_bounds}.  These constraints ensure that no process conditions, environmental regulations, toxicity thresholds, etc. are violated with the designed structures.

Mixture design problems or single component design problems involving mixtures typically have a number of common constraints.  Most notably, these problems often require some idea of the activities of the chemical species in the mixture and thus necessitate incorporating an activity coefficient model.  There are three such models often used in CAMD:
\begin{enumerate}
\item \textbf{UNIFAC.}  UNIFAC~\cite{fjp75} is a group contribution variant of the UNIQUAC equation~\cite{ap75} and has been used extensively in CAMD.  UNIFAC is simple, accurate, and easy to incorporate into CAMD problems due to its use of groups.  The original UNIFAC may lack accuracy outside of standard temperature and pressure ranges, but many extensions and re-parameterizations exist to address these cases.
\item \textbf{SAFT.} SAFT~\cite{saft} is an accurate equation of state that is applicable in many temperature and pressure domains.  It is gaining popularity in CAMD and several group contribution methods~\cite{tkvmc07,lagj07,lajg08,pgdm09,plaajg14} have already been developed to estimate the SAFT parameters necessary for its use.
\item \textbf{COSMO-RS and -SAC}  COSMO-RS~\cite{k95,kvbl98} and COSMO-SAC~\cite{ls02iecr} are two post-processing methods for the COSMO solvation model~\cite{ks93}.  Unlike other approaches, these COSMO-based models use electronic surface charge distributions that are calculated at the quantum chemistry level.  These methods are now more amenable to CAMD due to the development of various COSMO-based group contribution methods~\cite{mrg07,mrg09,ast16}.
\end{enumerate}

\subsection{\mbox{Forms of the structural feasibility} \mbox{constraints}}

The constraints defined above in \eqref{camd_prop_ineq} and~\eqref{camd_prop_eq} ensure that the structural descriptors chosen with the variables $n_d$ are consistent with all other structural features such that they can be assembled to create a chemically feasible molecule.  The three different types of QSPRs given above require different sets of these constraints, so this discussion will be divided into three parts, beginning with group contribution methods.

\subsubsection{\mbox{Connectivity constraints for GC} \mbox{methods}}

\begin{figure*}
\footnotesize
\caption{An example of connectivity constraints for GC methods, considering an acyclic, aliphatic molecule\label{struct_constraints_gc}}
\centering
\begin{tabular}{cc}
\begin{tabular}{c}
\begin{tabular}{|c|cccc|c|}
\hline
Groups &
\adjustbox{valign=c}{\includegraphics[scale=1.0]{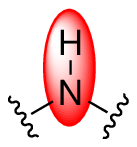}} &
\adjustbox{valign=c}{\includegraphics[scale=1.0]{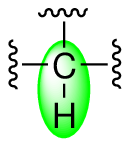}} &
\adjustbox{valign=c}{\includegraphics[scale=1.0]{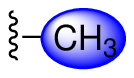}} &
\adjustbox{valign=c}{\includegraphics[scale=1.0]{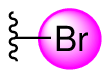}}  \\
\hline
Valence ($\Phi^{ali}_d$) & 2 & 3 & 1 & 1 \\
\hline
$n_d$  & 1 & 1 & 2 & 1 \\
\hline
$n_d(2-\Phi^{ali}_d)$ & 0 & -1 & 2 & 1 \\
\hline
\end{tabular}
\\
\begin{tabular}{C{2cm}c}
Example molecule &
\adjustbox{valign=c}{\includegraphics[scale=1.0]{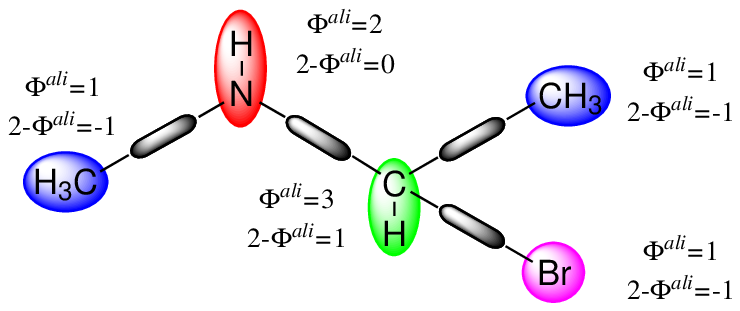}}\\
\end{tabular}
\end{tabular}
&
\begin{tabular}{|c|}
\hline
Feasibility check \\
\hline
\textit{Valence balance constraints} \\
$\sum\limits_{d} (\Phi^{ali}_d-2)n_d = 2 $\\
$ \stackrel{?}{=} 2-2(N_R^\text{ali}) =2 \, \checkmark $\\
\\
\textit{Sufficient number of groups} \\
\textit{for satisfying valence} \\
$\sum\limits_d n_d = 5$\\
$ \stackrel{?}{\geq} n_{\text{NH}_\text{1}}(\Phi_{\text{NH}_\text{1}}-1)+2 = 3 \, \checkmark $ \\
$ \stackrel{?}{\geq} n_{\text{CH}_\text{1}}(\Phi_{\text{CH}_\text{1}}-1)+2 = 4 \, \checkmark $ \\
$ \stackrel{?}{\geq} n_{\text{CH}_\text{3}}(\Phi_{\text{CH}_\text{3}}-1)+2 = 2 \, \checkmark $ \\
$ \stackrel{?}{\geq} n_{\text{Br}}(\Phi_{\text{Br}}-1)+2 = 2 \, \checkmark $ \\
\hline
\end{tabular}
\end{tabular}
\end{figure*}

If using GC methods to solve CAMD problems, the variable $n_d$ typically represents the number of occurrences of the group $d$ in the solution.  Many structural constraints in CAMD also consider the valence of each group, $\Phi_d$, which is simply the number of bonds a group requires to satisfy its valence electron requirements.  For example, the group $-\text{CH}_\text{1}\text{(Cl)}-$ requires 2 external bonds ($\Phi_{\text{CH}_\text{1}\text{(Cl)}}=2$) as two of carbon's four required bonds are accounted for by hydrogen and chlorine atoms.  One of the most widely-used rules for structural feasibility using valences comes from Odele and Macchietto~\cite{om93}:
\begin{align}
\sum\limits_d (\Phi_d-2)n_d = 2m
\end{align}
where m is defined by:
\begin{align}
m=
\begin{cases}
    -1 ,& \text{if compound is acyclic} \\
    0 ,& \text{if compound is monocyclic} \\
    1 ,& \text{if compound is bicyclic}
\end{cases}
\end{align}
These constraints alone, while appropriate for most molecules, still allow for certain sets of groups which cannot be joined to form feasible molecules.  For example, a solution of 1 $>\text{CH}_\text{0}<$ ($\Phi_{\text{CH}_\text{0}}=4$) group and 2 -Br ($\Phi_{\text{Br}}=1$) groups is feasible for the above constraints with $m=0$.  To address these situations, Odele and Macchietto also define a constraint to ensure that there are enough groups to meet the valence requirements of every group.  This means that to include a group in the solution with a valence requirement of 4, there must also be at least 4 additional groups.  This is captured in
\begin{align}
\sum\limits_d n_d \geq n_{d'}(\Phi_{d'}-1)+2 \quad \forall d'
\end{align}
where $d'$ is also an index over descriptors.  These constraints can be generalized with integer variables $N_R^\text{arom}$ and $N_R^\text{ali}$, which represent the number of aromatic rings and the number of aliphatic rings, respectively.  Furthermore, we introduce $G_{ali}$ and $G_{arom}$ to be the sets of groups which have any open valences which are aliphatic and aromatic, respectively.  We define these valence counts to be $\Phi^{ali}$ for aliphatics and $\Phi^{arom}$ for aromatics.  Note that the same group can appear in both sets.  We also define the parameter $A^\text{arom}_d$ to represent the number of aromatic atoms in a group $d$ and the parameter $\rho_d$ to represent the number of available aliphatic attachment points for every aromatic atom in group $d$.  Below we provide a slightly more general formulation considering aromatic rings:
\begin{align}
\sum\limits_{d \in G_{ali}} (2-\Phi^{ali}_d)n_d = 2 - 2N_R^\text{ali} \nonumber \\
+2\sum\limits_{d\in G_{arom}} \rho_d n_d-2N_R^{arom} \\
\sum\limits_{d \in G_{arom}} (2-\Phi^{arom}_d) n_d = 0  \\
\sum\limits_{d \in G_{arom}} A^\text{arom}_d = 6N_R^\text{arom} \\
\sum\limits_{d \in G_{ali}} n_d \geq n_{d'}(\Phi^{ali}_{d'}-1)+2 \quad \forall d'\\
\sum\limits_{d \in G_{arom}} n_d \geq n_{d'}(\Phi^{arom}_{d'}-1)+2 \quad \forall d'
\end{align}
We note that in this example, the aromatic rings are assumed to be benzylic and not attached to another aromatic ring (biphenyls and fused aromatics).  An example of these basic structural constraints for GC methods are given in Fig.~\ref{struct_constraints_gc}.  A few simple extensions allow these constraints to account for all aromatic rings.  These modified Odele-Macchietto constraints work in cases where all groups are bonded to each other with only single or aromatic bonds.  More complex connectivity constraints, accounting for cases where groups may be double or triple bonded to one another are given in~\cite{sty03}.

\subsubsection{\mbox{Connectivity constraints for TI} \mbox{methods}}

\begin{figure*}
\footnotesize
\caption{An example of connectivity constraints for TI-based CAMD\label{struct_constraint_ti}}
\centering
\makebox[\textwidth][c]{
\begin{tabular}{ccc}
Molecule & $a_{v,v',b=1}$ & $a_{v,v',b=2}$ \\
\hline
\adjustbox{valign=c}{\includegraphics[scale=1.2]{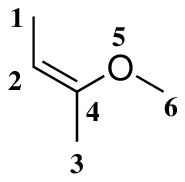}}
&
$
  \kbordermatrix{
    & v_1' & v_2' & v_3' & v_4' & v_5' & v_6' \\
    v_1 & 0 & 1 & 0 &  \multicolumn{1}{c}{\cellcolor{blue!20}0} & 0& 0 \\
    v_2 & 1 & 0 & 0 &  \multicolumn{1}{c}{\cellcolor{blue!20}0} & 0& 0 \\
    v_3 & 0 & 0 & 0 &  \multicolumn{1}{c}{\cellcolor{blue!20}1} & 0 & 0\\
    v_4 & 0 & 0 & 1 & \multicolumn{1}{c}{\cellcolor{red!20}0} &  \multicolumn{1}{c}{\cellcolor{blue!20}1} &  \multicolumn{1}{c}{\cellcolor{blue!20}0}\\
    v_5 & 0 & 0 & 0 & 1 & 0 & 1\\
    v_6 & 0 & 0 & 0 &0 & 1& 0
  }
  $
  &
  $
  \kbordermatrix{
    & v_1' & v_2' & v_3' & v_4' & v_5' & v_6' \\
    v_1 & 0 & 0 & 0 & \multicolumn{1}{c}{\cellcolor{green!20}0} & 0& 0 \\
    v_2 & 0 & 0 & 0 & \multicolumn{1}{c}{\cellcolor{green!20}1} & 0& 0 \\
    v_3 & 0 & 0 & 0 & \multicolumn{1}{c}{\cellcolor{green!20}0} & 0 & 0\\
    v_4 & 0 & 1 & 0 & \multicolumn{1}{c}{\cellcolor{red!20}0} & \multicolumn{1}{c}{\cellcolor{green!20}0} & \multicolumn{1}{c}{\cellcolor{green!20}0}\\
    v_5 & 0 & 0 & 0 & 0 & 0 & 0\\
    v_6 & 0  & 0 & 0 & 0 & 0& 0
  }
  $
  \vspace{3mm}
\\
\hline
\multicolumn{3}{c}{
Valence constraint for $v_4$
} \\
\hline
Valence requirements & Single bonds & Double bonds \\
\begin{tabular}{c}
$\Phi_{l=(\text{=CH}_\text{0}<),b=1}=2$ \\
$\Phi_{l=(\text{=CH}_\text{0}<),b=2}=1$
\end{tabular}
&
\colorbox{blue!20!white}{
\begin{tabular}{c}
$\sum\limits_{v'=1}^{v-1} a_{v',4,1}
+ \sum\limits_{v'=v+1}^{V} a_{4,v',1} = 1+1 $\\
$
 \stackrel{?}{=} (\Phi_{(\text{=CH}_\text{0}<),1})( y_{4,(\text{=CH}_\text{0}<)}) =2  \, \checkmark$
\end{tabular}
}
&
\colorbox{green!20!white}{
\begin{tabular}{c}
$\sum\limits_{v'=1}^{v-1} a_{v',4,2}
+ \sum\limits_{v'=v+1}^{V} a_{4,v',2} = 1+0 $\\
$
 \stackrel{?}{=} (\Phi_{(\text{=CH}_\text{0}<),2})( y_{4,(\text{=CH}_\text{0}<)}) =1  \, \checkmark$
\end{tabular}
}
\\
\hline
\multicolumn{3}{c}{
Connected graph constraint for $v'_4$
} \\
\hline
Vertex in solution? &\multicolumn{2}{c}{Sum of connections to lower-indexed vertices} \\
\begin{tabular}{c}
$\sum\limits_l y_{4,l \neq \text{dummy}} =1 $\\
vertex appears in\\
the solution (non-dummy)
\end{tabular}
&
\multicolumn{2}{c}{
\begin{tabular}{c}
$\sum\limits_{v=1}^{4-1} \sum\limits_{b} a_{v,v',b}
=\sum\limits_{v=1}^{3} a_{v,v',b=1}+\sum\limits_{v=1}^{3} a_{v,v',b=2}=\colorbox{blue!20!white}{0+0+1}
+\colorbox{green!20!white}{0+1+0}=2$ \\
$2 \stackrel{?}{\geq} 1 \checkmark $
\end{tabular}
}
\\
\hline
\end{tabular}
}
\end{figure*}

Constraints for TI-based CAMD work in a similar fashion to constraints for GC-based methods.  Many CAMD approaches using TIs do not account for topological features explicitly, meaning that there are usually no variables signifying the number of paths, Randi\'{c} indices, etc.  These values are instead calculated as a function of the molecular graph, which is typically represented in the form of an adjacency matrix $A$.  One of the most general adjacency matrices for TI applications is defined by Camarda and Maranas~\cite{cm99}.  The matrix they defined is 3-dimensional with dimensions corresponding to all of the nodes in the molecular graph on two axes and all the possible bond types on another axis.  The possible bond types given by Camarda and Maranas are $b=\{1,2,3\}$ for single, double, and triple bonds, respectively.  An entry, $a_{v,v',b}$, of the adjacency matrix is defined by:
\begin{align}
a_{v,v',b}=
\begin{cases}
1 ,& \begin{array}{l}
       \text{if vertex $v$ is connected to vertex $v'$} \\
       \text{with a bond of type $b$}
     \end{array} \\
    0 ,& \text{otherwise}
\end{cases}
\end{align}
Furthermore, there are assumed to be $L$ different types of nodes, where each of these nodes $l$ has a valence value $\delta_l$.  Camarda and Maranas appeal to the formulation of Raman and Maranas~\cite{rm98} and define a binary variable $y_{v,l}$ to be
\begin{align}
y_{v,l}=
\begin{cases}
1 ,& \text{if vertex $v$ is a node of type $l$} \\
    0 ,& \text{otherwise}
\end{cases}
\end{align}
Using these definitions, they define a value $\Delta_v$ which assigns the correct valence value based on the type of node:
\begin{align}
\Delta_v=\sum\limits_l \delta_l y_{v,l}
\end{align}
The 0th-order connectivity index can then be defined:
\begin{align}
^0\chi=\sum\limits_v \sum\limits_l \frac{y_{v,l}}{\sqrt{\Delta_v}}
\end{align}
The 1st-order connectivity index is defined similarly:
\begin{align}
^1\chi=\sum\limits_v \sum\limits_l \frac{\sum\limits_l a_{v,v',l}}{\sqrt{\Delta_v \Delta_{v'}}}
\end{align}
Other TIs calculated in this fashion are provided in~\cite{rm98,cm99,cs05} as well as a linearization of these relationships.

Finally, the structural feasibility constraints using this formulation begin first with an assignment constraint to ensure that each available vertex can only be assigned to one type:
\begin{align}
\sum\limits_l y_{v,l} =1 \quad \forall v
\end{align}
Note that not all of the vertices are used.  The Camarda and Maranas formulation defines ``dummy'' atom types, so that the optimal molecule may have fewer than the number of allowed vertices.  Additionally, only one bond type is possible between any two vertices $v$ and $v'$.  This is captured in:
\begin{align}
\sum\limits_b a_{v,v',b} \leq 1 \quad \forall v,v'
\end{align}
The valence-balance constraints again uses $\Phi_{l,b}$, which now signifies the number of external bonds of type $b$ required by a node of type $l$.  This constraint ensures that if a vertex $v$ requires two single bonds, there are exactly two $a_{v,v',b=1}$ which are equal to 1.  Note that the constraint is written so that only the upper right triangle of the connectivity matrix is needed.  An equivalent constraint is used for $b=2$.  These constraints are provided below:
\begin{align}
\sum\limits_{v'=1}^{v-1} a_{v',v,b}
+ \sum\limits_{v'=v+1}^{V} a_{v,v',b}
 = \sum\limits_l \Phi_{l,b} y_{v,l} \quad \forall v,b
\end{align}
Finally, a constraint is used to enforce that all vertices are connected.  Specifically, this constraint ensures that for any (non-dummy) vertex $v'$, there is at least one attached vertex with a smaller index.  In this way, the formation of two disjoint subgraphs is prevented.  The constraint is provided below:
\begin{align}
\sum\limits_{v=1}^{v'-1} \sum\limits_{b} a_{v,v',b}
\geq \sum\limits_l y_{v',l \neq \text{dummy}}  \quad \forall v' \in \{2,\hdots,V\}
\end{align}
The above captures the majority of necessary constraints for a CAMD problem.  Others are given in~\cite{cm99}.  We provide an example of these structural feasibility constraints in Fig.~\ref{struct_constraint_ti}.

\subsubsection{\mbox{Connectivity constraints for SD} \mbox{methods}}

Though several approaches to these constraints exist, perhaps the most intuitive for CAMD problems comes from Chemmangattuvalappil et al.~\cite{csbe10}.  These constraints are more complicated than in previous cases due to the fact that atomic signature descriptors are intended to overlap.  The main difficulty arises from the need for consistency among descriptors.  For example, the simple molecule of propane should have a height-1 atomic signature of $\text{C}^\text{2}\text{(C}^\text{1}\text{)(C}^\text{1}\text{)}$ which corresponds to the central atom.  In this case, the coloring scheme (represented by the superscripts) indicates how many non-hydrogen bonds an atom has.  Given the occurrence of one height-1 atomic signature of $\text{C}^\text{2}\text{(C}^\text{1}\text{)(C}^\text{1}\text{)}$, we would also expect two atomic signatures of the form $\text{C}^\text{1}\text{(C}^\text{2}\text{)}$ to exist.  These would correspond to the height-1 signatures of the terminal carbons.  Furthermore, the existence of the height-1 signature $\text{C}^\text{2}\text{(C}^\text{1}\text{)(C}^\text{1}\text{)}$ also requires the presence of three height-0 signatures corresponding to the atoms that make it up: $\text{C}^\text{1}$, $\text{C}^\text{1}$, and $\text{C}^\text{2}$.  This is not always necessary as some QSPRs using SDs only use atomic signatures of a certain height.  In cases where multiple heights are used, this consistency among different heights must be considered.

First, Chemmangattuvalappil et al.~\cite{csbe10} define an index $d$ over the set of all of the atomic signature descriptors, $D$.  There are also different subsets of atomic signature descriptors corresponding to atomic signatures whose parent node forms one external bond ($S_1$), two external bonds ($S_2$), three external bonds ($S_3$), and four external bonds ($S_4$).  Furthermore, there are subsets to define signatures whose parent node has one double bond ($B_{D1}$), two double bonds ($B_{D2}$), and one triple bond ($B_{T1}$).  To ensure that the graph produced is feasible, the authors use the \textit{handshaking lemma}, the graph property that the sum of the degrees of every node is equal to twice the number of edges ($\sum\limits_v \delta(v) = 2|E|$).  From graph theory, we also know that the number of edges in a connected graph is equal to the number of vertices minus 1 plus the dimension of the cycle space, $|R|$, which is the number of rings containing no other rings.  $|R|$ is usually intuitive from a chemical perspective and usually simply means the number of rings (e.g. naphthalene has 2 rings because the outer cycle of all carbons does not count, and norbornane has two rings because the cyclohexane ring is bridged and does not count).  In equation form, this relationship is $|E|=|V|-1+|R|$.  This implies that $\sum\limits_v \delta(v) = 2(|V|-1+|R|)$ and the first feasibility constraint follows:
\begin{equation}
\begin{aligned}
\sum\limits_{d \in S_1} n_d + 2\sum\limits_{d \in S_2} n_d +
3\sum\limits_{d \in S_3} n_d + 4\sum\limits_{d \in S_4} n_d \\
=2 \Bigg[ \left[ \sum\limits_{d \in D} n_d + \frac{1}{2} \sum\limits_{d \in B_{D1}} n_d
+\sum\limits_{d \in B_{D2}} n_d+\sum\limits_{d \in B_{T1}} n_d \right] \\
 -1 +N_R^\text{arom} +N_R^\text{ali} \Bigg]
\end{aligned}
\end{equation}
We note that this feasibility constraint only requires checking signatures of a single height.  In other words, if SDs of heights 0, 1, and 2 are used, it is only necessary to check this constraint for SDs of one (say, height-1) of these heights.

\begin{figure*}[ht]
\footnotesize
\centering
\caption{An example of structural constraints using SDs for a simple alkane\label{struct_constraint_sd}}
\begin{tabular}{C{5cm}|C{2.3cm}C{2.3cm}C{2.3cm}C{2.3cm}}
Original Structure & \multicolumn{4}{c}{Atomic signature descriptors (height-1)} \\
\hline
\adjustbox{valign=c}{\includegraphics[scale=1.0]{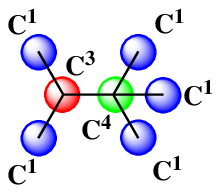}}&
\adjustbox{valign=c}{\includegraphics[scale=1.0]{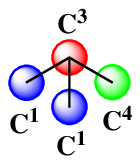}} &
\adjustbox{valign=c}{\includegraphics[scale=1.0]{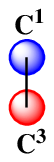}} &
\adjustbox{valign=c}{\includegraphics[scale=1.0]{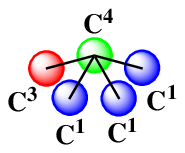}} &
\adjustbox{valign=c}{\includegraphics[scale=1.0]{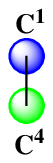}} \\
\hline
Number of occurrences ($n_d$) & 1 & 2 & 1& 3 \\
\hline
Degree count of parent atom ($\delta(v)$) : & 3 & 1 & 4& 1 \\
\hline
Number of coloring sequences ($\beta_{d,h=1,o,o'}$) : \\
\adjustbox{valign=c}{\includegraphics[scale=1.0]{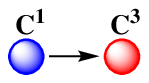}}& 0& 1& 0& 0\\
\adjustbox{valign=c}{\includegraphics[scale=1.0]{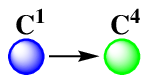}}& 0& 0& 0& 1\\
\adjustbox{valign=c}{\includegraphics[scale=1.0]{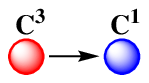}}& 2& 0& 0& 0\\
\adjustbox{valign=c}{\includegraphics[scale=1.0]{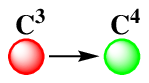}}& 1& 0& 0& 0\\
\adjustbox{valign=c}{\includegraphics[scale=1.0]{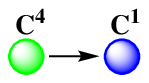}}& 0& 0& 3& 0\\
\adjustbox{valign=c}{\includegraphics[scale=1.0]{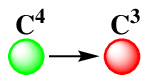}}& 0& 0& 1& 0\\
\hline
\multicolumn{5}{c}{Checking structural feasibility constraints}
\\
\hline
\multicolumn{5}{c}{
\begin{tabular}{cccc}
\begin{tabular}{c}
$\sum\limits_d \beta_{d,h,o,o'}n_d = \sum\limits_d \beta_{d,h,o',o} n_d$
\end{tabular} &
\begin{tabular}{c}
$1 \rightarrow 3$ and $3 \rightarrow 1$ \\
$(1)(2) = (2)(1) \, \checkmark $
\end{tabular}
&
\begin{tabular}{c}
$1 \rightarrow 4$ and $4 \rightarrow 1$ \\
$(1)(3) = (3)(1) \, \checkmark $
\end{tabular}
&
\begin{tabular}{c}
$3 \rightarrow 4$ and $4 \rightarrow 3$ \\
$(1)(1) = (1)(1) \, \checkmark $
\end{tabular}\\
\end{tabular}
} \\
\hline
\multicolumn{5}{c}{
\begin{tabular}{c}
$\sum\limits_{d \in S_1} n_d + 2\sum\limits_{d \in S_2} n_d +
3\sum\limits_{d \in S_3} n_d + 4\sum\limits_{d \in S_4} n_d $\\
$=2 \left[ \left[ \sum\limits_{d \in D} n_d + \frac{1}{2} \sum\limits_{d \in B_{D1}} n_d
+\sum\limits_{d \in B_{D2}} n_d+\sum\limits_{d \in B_{T1}} n_d \right] -1 +N_R^\text{arom} +N_R^\text{ali} \right]$ \\
$ 5 + 2(0) +
3(1) + 4(1)
=2 \left[ \left[ 7 + \frac{1}{2} (0)
+0+0 \right] -1 +0 +0 \right] \quad \iff \quad 12=12 \checkmark$
\end{tabular}}\\
\hline
\end{tabular}
\end{figure*}

Next, the authors define a ``coloring sequence'' to be a directed connection from a parent node to one of its children.  For example, propane's central atom would have a coloring sequence of $\text{C}^\text{2} \rightarrow \text{C}^\text{1}$ which represents the connection from the middle atom to one of the terminal carbons.  To produce a set of equivalent signatures, we would expect to find as many $\text{C}^\text{1} \rightarrow \text{C}^\text{2}$ coloring sequences as there are $\text{C}^\text{2} \rightarrow \text{C}^\text{1}$ coloring sequences.  This must be true for every signature height.  These considerations can be formulated as linear constraints using the variables $n$.  Considering $\beta_{d,h,o,o'}$ to be the number of coloring sequences $o \rightarrow o'$ for an atomic signature descriptor $d$ and height $h$, we can define the constraint:
\begin{align}
\sum\limits_d \beta_{d,h,o,o'}n_d = \sum\limits_d \beta_{d,h,o',o} n_d \quad \forall o,o',h
\end{align}
For coloring sequences that go from one coloring to an identical coloring, we expect the total number of these color sequences must be divisible by two.  More formally, for a color sequence which goes from coloring $o$ to coloring $o$ ($o \rightarrow o$), the following must apply:
\begin{align}
\sum\limits_d \beta_{d,h,o,o} n_d = 2K   \quad \forall h,o
\end{align}
where $K$ is a non-negative integer.  Finally, the authors introduce a constraint to ensure the connectivity of the graph.  For an atomic signature at a certain height $>0$, there must be a root (parent) atom and children atoms.  In some cases, there will be multiple children of the same coloring.  To produce a connectable, consistent set of signature descriptors, each of those children atoms must appear as a root atom in another signature.  This is captured by enforcing there to be at least as many signatures with a certain coloring of root (parent) atom as there are atoms with that coloring in child nodes.  In constraint form, this is:
\begin{align}
\sum\limits_{d \in C_{o' \rightarrow o}} n_d \omega_{d,o} \leq \sum\limits_{d \in S_{o \rightarrow o'}} n_d
\end{align}
The set $C_{o' \rightarrow o}$ is the set of all signatures who have parent nodes of coloring $o'$ and multiple ($\omega_{d,o}>1$) child nodes of coloring $o$.  Additionally, $\omega_{d,o}$ only includes vertices which have a degree greater than or equal to their parents.  $S_{o \rightarrow o'}$ is the set of signatures with a parent to child coloring sequence of $o \rightarrow o'$.  Additional information on this approach can be found in~\cite{csbe10}.  Alternative approaches to the constraints and the CAMD problem using SDs are given in~\cite{crmvklsbf04} and~\cite{wv10}.  A graphical example of some of these constraints is provided for an example in Fig.~\ref{struct_constraint_sd}.

\section{Techniques for solving the molecular design problem}
\label{camd_techniques}
\subsection{Generate-and-test methods}

Many QSPRs, especially those often used in CAMD, are simple functions which require little computational effort to evaluate.  Methods such as those discussed above are able to provide property estimates for millions of structures in a matter of minutes.  For this reason, many CAMD approaches have applied QSPR models in the ``forward'' direction, generating a large number of candidate structures and then evaluating every property of interest for every molecule.  This approach---known as ``generate-and-test''---has its merits in that it can often optimize over a pool of molecules without solving a potentially difficult optimization problem.  This is especially advantageous if the pool of potential molecules has been reduced to a practical number, either by considering a small problem or by one of many knowledge-based reduction procedures~\cite{cgn11,hg00,hg99}.  Of course, problems with a large design space are not solved efficiently with this approach.  In these cases, optimization methods have a distinct advantage over generate-and-test procedures.

Generate-and-test algorithms are fairly intuitive.  The primary requirements are consideration of every possible structure and low redundancy.  To that effect, several generate-and-test algorithms exist for groups~\cite{hg99,joback89}, topological indices~\cite{khf93,hdk93}, and signature descriptors~\cite{fcv03}.

\subsection{Decomposition methods}

Many CAMD problems are characterized by a large combinatorial space of potential descriptors, challenging non-linearities in thermodynamic of process models, and/or high sensitivity of the objective to process and descriptor variables.  As a result, many CAMD problems are too difficult to be addressed directly as optimization problems and  must instead be solved as a series of optimization subproblems.  Typically, these subproblems successively apply increasingly difficult constraints from the original problem, reducing the feasible set of molecules upon the solution of each subproblem.  This step-wise reduction in problem space makes many CAMD problems significantly easier.  Alternatively, many decomposition approaches approximate a set of constraints with a lower-bounding surrogate (in the case of minimization), devise a subproblem to generate feasible points, and then iterate between these subproblems until the upper and lower bounds on the objective function are sufficiently close.  Decomposition techniques can be broadly divided into the three categories of CAMD problems to which they are most often applied.  These are discussed below.

\subsubsection{\mbox{Decomposition in single molecule} \mbox{design}}

The most common technique in single molecule design involves systematically reducing the space of feasible molecules by applying increasingly stringent constraints.  For example, referring to the molecular design formulation shown in the previous section, some techniques (e.g.~\cite{hg99}) first apply the structural feasibility constraints, \eqref{camd_prop_eq} and~\eqref{camd_prop_ineq}, resulting in a set of all structurally feasible $n$ vectors.  This set of $n$ vectors represents a significantly smaller feasible region than that of the entire space defined by the descriptors.  These $n$ vectors can be assembled into feasible structures, and this set of structures can be evaluated one-by-one based on the remaining constraints and objective function or used in another optimization problem.

This approach works best if the number of feasible structures is reduced to a reasonably small number before difficult constraints are evaluated or the remaining optimization subproblem is solved.  For this reason, decomposition methods have seen the most success in cases where there are few possible descriptors, the constraints are very tight, or the problem involves minimizing a distance to property targets.  In cases where many $n$ vectors are possible, decomposition methods should be paired with optimization methods rather than generate-and-test methods.

\subsubsection{Decomposition in mixture design}

One of the more challenging problems in CAMD, the mixture design problem benefits from decomposition.  Early work from Gani and Fredenslund~\cite{gf93} proposed a decomposition algorithm based on the prior work of Klein et al.~\cite{kwg92} to decouple the mixture design problem into several single-component molecular design problems.  Each of these solutions could then be investigated as a potential mixture component.  Many other approaches have followed suit~\cite{kag05,cgn11,asst16}, relying on the efficiency of single-molecule CAMD techniques to quickly solve single-molecule subproblems and optimization techniques to optimize over the space of the mole fractions.

\subsubsection{\mbox{Decomposition in integrated} \mbox{product/process design}}

Integrated product/process design problems are also often decomposed.  A popular method relies on optimizing the process variables while allowing for any possible properties of the products.  This is a relaxation of the original problem and represents a lower bound (in the case of minimization) on the objective.  Solving this problem is far easier and generates a set of ideal properties for an optimal molecule to have.  Using these ideal properties as targets, a molecular design problem can be solved to determine the molecule that is closest in properties to these ideal values.  Finally, the process design problem can be updated with actual property values of the closest structure and then re-optimized.  This technique always produces a feasible value of the objective, though it may not always find the globally optimal structure(s) and process variable values.  This two-stage approach has been used by Eden et al.~\cite{ejge04} and Bardow et al.~\cite{bsg10}.  The latter reference has used this approach throughout the literature, referring to it as the continuous molecular targeting approach (CoMT-CAMD).

Other approaches optimize first in the space of the molecular structures before investigating process performance.  Papadopoulos and Linke~\cite{pl05} designed molecules based on a multi-objective optimization problem, determining the best potential structures based on a number of criteria.  This resulted in a Pareto-optimal set of structures, each of which could be tested in the context of the process design problem.  Approaches similar in nature~\cite{kag05,asst16} have also optimized in the space of structures and evaluated each structure/mixture as an input to process design problems.

Other approaches solve the problem iteratively.  For example, Buxton et al.~\cite{blp99} proposed iteratively solving two subproblems: one to identify process conditions and the other to determine molecular structures.   A new approach from Gopinath et al.~\cite{gjga16} uses the outer approximation algorithm~\cite{dg86} to treat the product/process design problems as two subproblems.  In this approach, one subproblem solved the process problem for a fixed molecule and the other determined another candidate molecule.  Approaches like these two discussed work by generating upper and lower bounds on the objective by solving the subproblems.  When the lower and upper bounds are within a certain tolerance, the algorithms terminate.

\subsection{Mathematical optimization methods}

In some cases, CAMD problems can be addressed straightforwardly with optimization techniques.  In others, optimization approaches become tenable with a slight alteration to the formulation or by exploiting the problem structure.  To reiterate an earlier statement, optimization approaches are best suited to problems with many possible descriptors or challenging non-linearities or non-convexities in molecular, mixture, or process models.  In cases where there are few possible descriptors, the design space can often be enumerated efficiently using generate-and-test methods.  Before discussing a few relevant techniques, we note that there is a large degree of overlap between the categories of decomposition methods and mathematical optimization methods.  Both techniques, however, merit an independent discussion as both play a critical role in the solution of CAMD problems.

Many techniques in single-molecule design are able to first solve feasibility optimization problems based on property and structural feasibility constraints.  This also qualifies as a decomposition method as the objective function is evaluated afterwards or structure ranking is presumed to be done by expert analysis.
Odele and Macchietto~\cite{om93} introduced a general optimization formulation for solving molecular design problems.  The problem discussed in their paper, the single molecule design problem, can be addressed very efficiently with optimization techniques, even if the descriptor space is large.

Duvedi and Achenie~\cite{da96} applied outer-approximation~\cite{dg86} to the formulation given by Odele and Macchieto with a few slight alterations.  A variant of this approach was also used by Churi and Achenie~\cite{ca96}, who considered connectivity of the structures.  In both cases, the problem was formulated as a mixed-integer nonlinear program (MINLP), which was solved with an outer approximation algorithm.  Though efficient for many types of problems, the outer approximation algorithms used in these cases are not guaranteed to find globally optimal solutions if the problem is non-convex.

Sinha et al.~\cite{sao99} and Sahinidis et al.~\cite{sty03} applied the branch-and-bound algorithm to these problems.  This required deriving underestimators (linear in many cases) of the constraints and objective function.  Solving the problem with underestimators resulted in a lower bound on the objective (in the case of minimization).  Then, variables in the problem corresponding to descriptor counts, physical property values, etc. would be divided into two regions, or ``branched on.''  This approach allowed optimization strategies to quickly disregard areas of the search space which would not lead to optimal solutions.  Furthermore, upon convergence, the branch-and-bound algorithm guarantees a globally optimal solution, at least to within a user-specified tolerance.

The problem of designing molecules to match property targets is far easier.  This problem can be solved in the mixed-integer quadratic program (MIQP) shown above or as a mixed-integer linear program (MILP), where some transformations of the QSPR functions must be done before optimization~\cite{sty03}.  This MILP has an altered objective function
\begin{align}
\underset{n,x}{\min} \sum\limits_k w_k(d_k^+ + d_k^-)
\end{align}
where $d_k^+$ and $d_k^-$ are the positive and negative deviations between the target property values and the estimated values.  Both $d_k^+$ and $d_k^-$ are modeled as positive continuous variables.  They require one additional constraint to calculate:
\begin{align}
d_k^+ - d_k^- = p_k^T - p_k \quad \forall k
\end{align}
Samudra and Sahinidis~\cite{ss13} used this observation to decompose the molecular design problem into three basic steps: (1) identification of optimal sets of groups, $n$; (2) structure generation; and (3) application of higher-order models to feasible structures.  Zhang et al.~\cite{zcg15} extended step (1) of this methodology to account for higher-order groups by introducing variables to account for the connectivity.  Maranas~\cite{mar96} proposed linearization of difficult property models to allow for the efficient use of MILP techniques.  Camarda and Maranas~\cite{cm99} convexified non-linear terms in their CAMD formulation, simplifying the location of globally optimal structures.

\subsection{Heuristics}

Though mathematical programming approaches are very useful, there are many CAMD problems which are too high-dimensional or non-linear to be practically considered by mathematical optimization approaches.  Problems of this type may attempt to design very large, structurally diverse molecules, necessitating a large search space with many possible combinations of descriptors.  Alternatively, these problems may involve a complex process design component, which may require a difficult simulation for each feasible structure.  Perhaps the thermodynamic functions and mole fractions optimization must be decoupled from the molecular design problem, meaning that each feasible structure must be investigated in a difficult mole fractions optimization subproblem.

Heuristic approaches to complex optimization problems typically apply high-level selection strategies to generate a series of trial points, evaluating the objective for each trial point and determining a new trial point based on the value of the objective as well as the history of function evaluations.  These algorithms terminate either by converging to a point of optimality or by exceeding some user-defined limit (time, iterations, etc.).  In the context of CAMD problems, most heuristic optimization approaches have optimized in the space of molecular structures.  Practically speaking, this involves generating a molecular structure based on certain design specifications (descriptors, property targets, etc.).  Each structure is fixed as an input to the entire CAMD problem, and the objective value is determined.  A new structure (or series of structures) is chosen for evaluation based on the relationship between structures/descriptors/properties and the objective value.

We note that one of the most important considerations for using heuristics for CAMD problems is the translation of molecular structures into a particular encoding.  In simple cases, the encoding can represent the number of descriptors used in the molecules.  In other cases, the descriptors may be translated into a binary string or assigned a value that is a function of the descriptors.

\subsubsection{Genetic algorithms}

Genetic algorithms (GAs) are a class of heuristic algorithms roughly based around the idea of natural selection.  More specifically, GAs evaluate the performance of each member of a certain pool of solutions called a ``generation.''  The members of the generation are then combined, taking some features from certain members---or ``parents''---and some from another.  The likelihood of being chosen for this ``reproduction'' process is based on each solution's rank, with preference given to better solutions.  In their application to CAMD problems, GAs typically work in the space of molecular structures.  Each structure in a generation is evaluated, features from the best-performing molecules are passed on to the next generation, and the process continues until convergence is achieved.

Venkatasubramanian et al.~\cite{vcc95} introduced GAs to CAMD problems.  They encoded molecular structures as a string of their substituent groups and applied genetic operations to the parent strings.  van Dyk and Nieuwoudt~\cite{vn00} proposed an encoding based on the UNIFAC~\cite{fjp75} groups.  Xu and Diwekar~\cite{xd05} developed a GC-based GA which encoded structures based on various group identities.  Herring and Eden~\cite{he15} applied GAs in conjunction with signature descriptors to design structures based on property targets.  Zhou et al.~\cite{zwms16} applied GAs to solvent design problems for a two-phase reactions and to process design problems for gas absorption~\cite{zzs16}.  Scheffczyk et al.~\cite{sfslb16} also applied GAs to liquid-liquid extraction problems based on COSMO-RS thermodynamics.

\subsubsection{Tabu search}

Tabu search (TS) algorithms also work by first proposing a pool of initial solutions.  These solutions are altered by one of many operations to produce slightly altered solutions.  This process is repeated so long as the altered molecules do not appear in a tabu list, i.e., a list of solutions forbidden from consideration based on various factors.  These factors can include: frequency of occurrence (to ensure the same solutions are not always visited), infeasibility or low-objective values (to allow the algorithm to determine good solutions), and many others.  Using this tabu list, the TS algorithm maintains some ``memory'' of previous solutions, which can offer some advantages in CAMD problems.  Naturally, the analogue of TS in CAMD problems maintains a solution set of molecular structures and alters these provided they are not on the tabu list of forbidden structures.  One important feature of TS algorithms is that the tabu list is often dynamic, meaning once tabu-forbidden solutions may be acceptable during a later generation.

Tabu search has only recently been applied to CAMD problems.  For example, Chavali et al.~\cite{clmc04} and Lin et al.~\cite{lc05} applied TS to the design of transition metal catalysts.  Another example comes from McLeese et al.~\cite{mehsc10}, who considered ionic liquid design.

\subsubsection{Other methods}

Several other heuristics have been applied to CAMD problems.  For example, Gebreslassie and Diwekar~\cite{gd15} solved liquid-liquid extraction problems with a modified ant colony optimization (ACO) algorithm.  Like other heuristics, ACO works by proposing a pool of solutions (molecules).  These solutions are assigned a certain weight, and good solutions attract other solutions in their direction, meaning that bad solutions become more like good solutions in properties, descriptors, or otherwise.  Similarly, bad solutions discourage other solutions from becoming similar to them.

Simulated annealing is another heuristic to produce solutions for the CAMD problem.  It works by altering a given solution by randomizing the descriptors that define its encoding.  If the new solution is better than the previous, it is accepted and becomes the current solution.  If the new solution is worse but still within the bounds of an error function, it is also accepted.  As the algorithm proceeds, the error function becomes more and more stringent, meaning that worse solutions are less likely to be accepted.  Examples of this algorithm in CAMD come from Ourique and Telles~\cite{ot98} and Marcoulaki and Kokossis~\cite{mk98}.

\section{Applications of CAMD}
\label{camd_applications}
\subsection{Single molecule design}

To begin, we mention a few approaches for single molecule design which qualify as solving the ``feasibility problem.''   As discussed earlier, it is sometimes difficult to make the distinction between this category of problems---feasibility problems---and those that have some ranking criteria.  The simple reason for this is that many methodologies are designed to do both.  Nonetheless, this type of application is presented here to underscore its importance.  A few noteworthy examples come from Joback~\cite{joback89} and Joback and Stephanopoulos~\cite{js90,js95}, who describe techniques to solve the feasibility problem using groups, and Kier and co-workers~\cite{kh93,khd93,khf93}, who solved this problem using topological indices, and Churchwell et al.~\cite{crmvklsbf04}, who described techniques to solve these types of problems with signature descriptors.  All of these studies, though not directly applying optimization, use many techniques to reduce the massive size of the chemical search space.  Many techniques described in subsequent sections can solve these feasibility problems with optimization approaches.  These methods typically have an explicit objective function and thus fit better into a different category.

The next category of applications concerns problems with explicit objective functions.  The many objective functions of this category reflect the diversity of the CAMD field as a whole.  Some of the first contributions in CAMD came from Gani and Brignole~\cite{gb83} and Brignole et al.~\cite{bbg86} who designed extraction solvents based on solvent power and selectivity.  Though Gani and Brignole did not use optimization in this approach, they significantly reduced the  number of feasible molecular structures based on arguments like limiting groups, placing restrictions on various properties, and investigating solubility curves.  In this way, they narrowed the massive chemical design space to a few possible molecules which could all be considered directly by solving the ``forward problem.''  A similar approach was used by Macchietto et al.~\cite{moo90} and Joback~\cite{joback89}.

Early efforts at optimization in the CAMD problem come from Gani et al.~\cite{gnf91} and  Knight and McRae~\cite{km91}.  Odele and Macchietto~\cite{om93} introduced a general numerical optimization to the CAMD problem, formulating several solvent design problems as optimization problems over the space of groups.  Since then, many CAMD efforts have focused on the design of solvents.  For example, Harper et al.~\cite{hg99} proposed a multi-level approach to design separations solvents, reducing the number of feasible molecules in each of several sub-problems.  Sinha et al.~\cite{sao99} applied global optimization techniques to the design of a blanket wash solvent.  Karunanithi et al.~\cite{kag05} developed a decomposition algorithm to address difficult optimization problems, applying it to the design of liquid-liquid extraction solvents and crystallization solvents~\cite{kag06}.  Subsequent investigations of this problem came from Samudra and Sahinidis~\cite{ss13} and Austin et al.~\cite{asst16}.

Many efforts in CAMD have been applied towards industrial processes.  One of the most popular application areas is in designing solvents for liquid-liquid extraction.  Gani et al.~\cite{gnf91} proposed the design of a solvent for the separation of water and acetic acid.  Harper et al.~\cite{hg99} and Harper and Gani~\cite{hg00} sought to find a replacement for toluene in the separation of phenol and water.  A few other examples include: Marcoulaki and Kokossis~\cite{mk98}, Xu and Diwekar~\cite{xd05}, Scheffczyk et al.~\cite{sfslb16}, and Gebreslassie and Diwekar~\cite{gd15}.

Extractive distillation is also another important industrial process which is often studied using CAMD.  Many approaches investigated focused on separations in general and have also been applied to liquid-liquid extraction.  Two of those such studies are Harper et al.~\cite{hg99} and Gani et al.~\cite{gnf91}.  van Dyk and Nieuwoudt~\cite{vn00} applied simulated annealing to the design of a solvent separation of five binary pairs of common industrial compounds.

Another emerging area of study in CAMD is the design of solvents to optimize reaction properties.   Wang and Achenie~\cite{wa02b} made one of the first efforts to solve this problem, designing solvents to promote ethanol fermentation and subsequent extraction.  Gani et al.~\cite{gjc05} proposed a rules-based strategy for the selection of solvents for several common reactions and a pharmaceutical example.  Foli\'{c} et al.~\cite{fap07,fap08} designed novel solvents to maximize the reaction rate constant of an $S_N2$ reaction. They applied GC methods to estimate the parameters of the the the solvatochromic equation of Abraham et al.~\cite{adk+87} and then used this equation to predict rate constants.   Struebing et al.~\cite{sgk+13} investigated the same type of reactions, now using quantum chemical calculations for their solvents and the solvatochromic equation as a surrogate model.  Zhou et al.~\cite{zlqs15} used COSMO-RS~\cite{k95} thermodynamics in conjunction with CAMD techniques to design solvents to maximize reaction selectivity~\cite{zlqs15}.  Austin et al.~\cite{ast16} designed solvents to maximize a reaction rate using COSMO-RS thermodynamics and projecting the original problem onto a lower-dimensional space.

Ionic liquid design is another application area for this subset of CAMD problems.  For example, Karunanithi and Mehrkesh~\cite{km13} used decomposition techniques to design ionic liquids for electrical conductivity, heat transfer, liquid-liquid separations, and solubility.  McLeese et al.~\cite{mehsc10} designed ionic liquids using a tabu search algorithm, which they also showed produced a globally optimal solution for their test problem.  Matsuda et al.~\cite{mykt07} designed ionic liquids based on conductivity and viscosity targets.  These case studies used a small number of possible groups and could thus be alternatively addressed via exhaustive enumeration.

A number of approaches have also investigated pharmaceutical applications of CAMD.  We note here that a good review exists on the topic of pharmaceutical solvents already~\cite{rha13}.  Chemmangattuvalappil et al.~\cite{csbe10} investigated a drug modification problem using molecular signatures.  Their approach, discussed above, required a fairly large number of linear constraints to ensure a consistent set of descriptors.  They applied this approach to the alteration of alkyl substituents on a fungicidal compound.  Siddhaye et al.~\cite{scst04} designed pharmaceutical products, focusing on molecules likely to be the active pharmaceutical ingredient (API).  Leveraging the power of connectivity indices, they were able to design a few pharmaceutically relevant case studies, including producing a penicillin derivative with specified properties.  Churchwell et al.~\cite{crmvklsbf04} used signature descriptors to design peptide inhibitors to leukocyte functional antigen-1 (LFA-1) and its ligand intercellular adhesion molecule-1 (ICAM-1).

Many CAMD efforts have also been focused on the design of alternative refrigerants.  For example, Gani et al.~\cite{gnf91} designed refrigerants based on a few important properties.  Joback~\cite{joback89} first considered the problem of finding a replacement refrigerant for Freon-12.  Duvedi and Achenie~\cite{da96,da97} and Churi and Achenie~\cite{ca96} looked at this same problem, focusing on heat capacity and heat of vaporization as the properties to optimize.  Marcoulaki and Kokossis~\cite{mk98} also designed a replacement for Freon-12 using a simulated annealing approach.  Sahinidis et al.~\cite{sty03} investigated the same replacement problem using a global optimization approach with modified structural constraints and an improved CAMD formulation.  Samudra and Sahinidis~\cite{ss13iecr} designed heat transfer fluids for refrigeration systems.

Polymer design has been investigated extensively using CAMD techniques.  Many of these approaches focused on designing a polymer with certain physical properties.  Venkatasubramanian et al.~\cite{vcc94,vcc95} applied genetic algorithms to the problem, designing polymers to approximate target values in various property categories like glass transition temperature, bulk modulus, heat capacity, density, and others.  Maranas~\cite{mar96}  approaches the problem in a similar way, minimizing distances to target values.  Unlike Venkatasubramanian et al., Maranas used a mixed-integer linear program (MILP) formulation and mathematical optimization techniques to solve the problem.  Camarda and Maranas~\cite{cm99} addressed the design problem using topological indices and a mathematical optimization formulation.  Eslick et al.~\cite{eyptsc09} also used topological indices but designed molecules with a tabu search algorithm rather than a mathematical optimization approach.  Brown et al.~\cite{bmrf06} considered the problem with signature descriptors.  Pavurala and Achenie~\cite{pa13} used an outer-approximation approach to design polymers to aid in oral drug delivery.

This section presents a selection of main applications of single-molecule design problems.  For a more comprehensive list, see Table~\ref{bigtable}.

\subsection{Mixture design}

The mixture design problem is a difficult variant of the single-molecule design problem.  As a result, there are far fewer examples of applications considering a mixed product.  Though we divide by application in this section, we emphasize that many of these techniques are generalizable.  Many of the references given here can likely be altered to design a mixed product for an arbitrary application.

Klein et al.~\cite{kwg92} and Gani and Fredenslund~\cite{gf93} first considered the mixture design problem, solving a few example problems involving solubilities and compounds which form azeotropes.  Vaidyanathan and El-Halwagi~\cite{re96} designed blends of polymers, relying on simple mixing rules for algebraic simplicity.  Vaidyanathan and El-Halwagi also designed single polymers.  Duvedi and Achenie~\cite{da97} developed an MINLP formulation for mixture design, using an equation of state to estimate some mixture properties and relative simple mixing rules to estimate others.  They applied the methodology to the design of refrigerant blends.

Buxton et al.~\cite{blp99} proposed a decomposition technique to solve mixture design problems.  They produced solvent blends which would reduce the environmental impact of an industrial process.  Sinha et al.~\cite{sag03} solved mixture selection problems as an MINLP, choosing the best combination of solvents from a given list.  This approach was able to select an optimal single solvent and mixture of solvents for use as cleaning agents in the lithographic printing industry.

Karunanithi et al.~\cite{kag06} used a decomposition technique to first reduce the mixture design problem to the set of all feasible individual components.  Each possible mixture was then used to evaluate the objective function.  This approach was used for the design of a crystallization solvent and anti-solvent.  Conte et al.~\cite{cgn11} proposed a task-based decomposition algorithm, which was applied to the design of paint solvents blends and insect repellent solvent blends.  Austin et al.~\cite{asst16} addressed the mixture design problem in the reduced-order space of each individual component's properties, employing derivative-free optimization methods to optimize over the lower-dimensional space.  This was applied to reproduce the crystallization solvent design problems of Karunanithi et al.~\cite{kag06}.

\begin{table*}
\caption{Summary of CAMD applications and the methodologies used in each case \label{bigtable}}
\begin{threeparttable}
\footnotesize
\begin{tabular}{L{5cm} L{10cm}}
\textbf{Application} & \textbf{References} \\
\hline
Antigen inhibition activity & Churchwell et al.~\mtnote{sd,d,o}~\cite{crmvklsbf04}  \\
\hline
Biodiesel additives & Hada et al.~\mtnote{gc,d,gt}~\cite{hse14} \\
\hline
$\text{CO}_\text{2}$ capture & Gani et al.~\mtnote{gc,d,gt,m}~\cite{gnf91}, Bardow et al.~\mtnote{sel,d,p}~\cite{bsg10}, Pereira et al.~\mtnote{gc,o,p}~\cite{pkg+11}, Stavrou et al.~\mtnote{sel,d,o,p}~\cite{slbg14}, Burger et al.~\mtnote{gc,o,p}~\cite{bpg+15}, Lampe et al.~\mtnote{gc,d,o,p}~\cite{lss+15}, Gopinath et al.~\mtnote{gc,d,o,p}~\cite{gjga16}   \\
\hline
Crystallization solvents & Karunanithi et al.~\mtnote{gc,d,gt,p,m}~\cite{kag06}, Samudra and Sahinidis~\mtnote{gc,d,o}~\cite{ss13}, Austin et al.\mtnote{gc,d,o,p,m}~\cite{asst16}  \\
\hline
Extractive distillation & Harper et al.~\mtnote{gc,d,gt}~\cite{hg99}, Gani et al.~\mtnote{gc,d,gt}~\cite{gnf91}, Papadopoulos and Linke~\mtnote{gc,d,h,p}~\cite{pl06b}, van Dyk and Nieuwoudt~\mtnote{gc,d,h}~\cite{vn00}\\
\hline
Extractive fermentation & Wang and Achenie~\mtnote{gc,d,o}~\cite{wa02b}, Papadopoulos and Linke~\mtnote{gc,d,h,p}~\cite{pl05} \\
\hline
Gas absorption & Odele and Macchietto~\mtnote{gc,d,o}~\cite{om93}, Buxton et al.~\mtnote{gc,d,o,m,p}~\cite{blp99}, Papadopoulos and Linke~\mtnote{gc,d,h,p}~\cite{pl06a,pl06b}, Bommareddy et al.~\mtnote{gc,d,gt,p}~\cite{bcse10}, Zhou et al.~\mtnote{gc,d,h,p}~\cite{zzs16}\\
\hline
HIV-1 protease inhibition activity & Visco et al.~\mtnote{sd,d,o}~\cite{vprf02} \\
\hline
Ionic liquids design & Matsuda et al.~\mtnote{gc,gt}~\cite{mykt07}, McLeese et al.~\mtnote{ti,h}~\cite{mehsc10}, Karunanithi and Mehrkesh~\mtnote{gc,d,h}~\cite{km13} \\
\hline
Liquid-liquid extraction   &    Gani and Brignole~\mtnote{gc,d,gt}~\cite{gb83}, Brignole et al.~\mtnote{gc,d,gt}~\cite{bbg86}, Odele and Macchietto~\mtnote{gc,d,o}~\cite{om93}, Marcoulaki and Kokossis~\mtnote{gc,d,h}~\cite{mk98}, Harper et al.~\mtnote{gc,d,gt}~\cite{hg99}, Harper and Gani~\mtnote{gc,d,gt}~\cite{hg00}, Gani et al.~\mtnote{gc,d,gt}~\cite{gnf91}, Karunanithi et al.~\mtnote{gc,d,gt,p}~\cite{kag05}, Austin et al.\mtnote{gc,d,o,p,m}~\cite{ast16}, Ourique and Telles~\mtnote{gc,h}~\cite{ot98}, Kim and Diwekar~\mtnote{gc,d,h,p}~\cite{kd02}, Papadopoulos and Linke~\mtnote{gc,d,h,p}~\cite{pl06a}, Xu and Diwekar~\mtnote{gc,d,h}~\cite{xd05}, Scheffczyk et al.~\mtnote{d,o}~\cite{sfslb16}, Gebreslassie and Diwekar~\mtnote{gc,h}~\cite{gd15}   \\
\hline
Organic Rankine cycle fluids & Papadopoulos et al.~\mtnote{gc,d,h,p}~\cite{psl10}, Lampe et al.~\mtnote{sel,d,p}~\cite{lsb+14}, Lampe et al.~\mtnote{gc,d,o,p}~\cite{lss+15} \\
\hline
Pharmaceutical products & Siddhaye et al.~\mtnote{ti,o}~\cite{scst04} \\
\hline
Polymer design & Venkatasubramanian et al.~\mtnote{gc,h}~\cite{vcc94,vcc95}, Maranas~\mtnote{gc,o}~\cite{mar96}, Vaidyanathan and El-Halwagi~\mtnote{gc,o,m}~\cite{re96}, Camarda and Maranas~\mtnote{ti,o}~\cite{cm99}, Brown et al.~\mtnote{sd,o}~\cite{bmrf06}, Eslick et al.~\mtnote{ti,h}~\cite{eyptsc09}, Pavurala and Achenie~\mtnote{gc,d,o}~\cite{pa13}, Zhang et al.~\mtnote{gc,o}~\cite{zcg15}\\
\hline
Reactions solvents  & Wang and Achenie~\mtnote{gc,d,o}~\cite{wa02b}, Gani et al.~\mtnote{sel,d}~\cite{gjc05}, Foli\'{c} et al.~\mtnote{gc,o}~\cite{fap07,fap08}, Struebing et al.~\mtnote{gc,d,o}~\cite{sgk+13}, Zhou et al.~\mtnote{gc,o}~\cite{zlqs15}, Austin et al.\mtnote{gc,d,o,m}~\cite{ast16}, Zhou et al.~\mtnote{gc,d,h}~\cite{zwms16} \\
\hline
Refrigerant design & Joback~\mtnote{gc,d,gt}~\cite{joback89}, Gani et al.~\mtnote{gc,d,gt}~\cite{gnf91}, Churi and Achenie~\mtnote{gc,d,o}~\cite{ca96}, Duvedi and Achenie~\mtnote{gc,d,o}~\cite{da96,da97}, Marcoulaki and Kokossis~\mtnote{gc,d,h}~\cite{mk98}, Sahinidis et al.~\mtnote{gc,o}~\cite{sty03}, Ourique and Telles~\mtnote{gc,h}~\cite{ot98}, Samudra and Sahinidis~\mtnote{gc,d,o}~\cite{ss13iecr}  \\
\hline
Separations (general) & Hostrup et al.~\mtnote{gc,sel,d,o,gt,p}~\cite{hhg99} \\
\hline
Solvents for consumer products and industry&  Pistikopoulos and Stefanis~\mtnote{gc,o}~\cite{ps98}, Buxton et al.~\mtnote{gc,d,o,m,p}~\cite{blp99}, Conte et al.~\mtnote{gc,d,gt,m}~\cite{cgn11}, Sinha et al.~\mtnote{gc,o}~\cite{sao99}, Sinha et al.~\mtnote{sel,o}~\cite{sag03}, Weis and Visco~\mtnote{sd,o}~\cite{wv10}\\
\hline
Soybean oil products & Camarda and Sunderesan~\mtnote{ti,o}~\cite{cs05} \\
\hline
Structural modifications to a fungicide & Raman and Maranas~\mtnote{ti,o}~\cite{rm98}, Chemmangattuvalappil et al.~\mtnote{sd,o}~\cite{csbe10} \\
\hline
Transition metal catalyst design & Chavali et al.~\mtnote{ti,h}~\cite{clmc04}, Lin et al.~\mtnote{ti,h}~\cite{lc05} \\
\hline
VOC recovery & Eden et al.~\mtnote{gc,d,gt,p}~\cite{ejge04} \\
\hline

\end{tabular}
\begin{tablenotes}
\footnotesize
\item[\textbf{gc}] Group contribution methods are the main QSPR method used
\item[\textbf{ti}] Topological indices are the main QSPR method used
\item[\textbf{sd}] Signature descriptors are the main QSPR method used
\item[\textbf{sel}] Compounds are selected from a fixed list rather than designed
\item[\textbf{d}] Decomposition methods used
\item[\textbf{gt}] Generate-and-test procedure used
\item[\textbf{o}] Numerical optimization used
\item[\textbf{h}] Heuristical optimization used
\item[\textbf{m}] Mixture design considered
\item[\textbf{p}] Process design considered
\end{tablenotes}
\end{threeparttable}
\end{table*}

\subsection{Integrated process and product design}

Though many CAMD endeavors design products with the ultimate goal of being incorporated into an industrial process, few have explicitly considered the relationship between a particular structure and a process.  This can be problematic as many recent efforts have observed some sensitivity between product descriptor variables and process variables.  To overcome this issue, various approaches have considered the product and process design problems simultaneously.

Eden et al.~\cite{ejge04} solved an integrated process and product design problem to best recover volatile organic compounds from an industrial process, identifying optimal property targets for solvents in a reduced-order space given by clustering methods.  Hostrup et al.~\cite{hhg99} proposed a general framework for integrated process and product design that was focused on separations.  This method relied on reduction of the feasible solution space via thermodynamic arguments and case-specific considerations.  Then, molecules were designed based on proximity to property targets for a certain process architecture.

Kim and Diwekar~\cite{kd02b} solved liquid-liquid extraction process problems, considering the process performance and designing suitable structures using a heuristic optimization strategy for the generation of solvent structures.  Papadopoulos and Linke~\cite{pl05} developed a methodology for considering integrated problems which relied on decomposing the problems into product and process subproblems. Unlike other approaches, Papadopoulos and Linke solved multi-objective optimization problems, determining the Pareto optimal front for solvent properties likely to be related to process performance.  Using these Pareto-optimal structures, they could solve the process problems for a much smaller set of possible molecules.  They applied this methodology to extractive fermentation~\cite{pl05} and liquid-liquid extraction and gas absorption~\cite{pl06a,pl06b}.

Karunanithi et al.~\cite{kag05} proposed a decomposition methodology to solve difficult process design problems.  This methodology first filtered out a large number of possible molecular structures based on property bounds.  It then applied a few stages of more complicated constraints to the remaining molecules, further reducing the pool of feasible structures.  Finally, the process model was applied to each of the molecules which were feasible for all of the constraints.  This methodology was applied to the design of a liquid-liquid extraction process~\cite{kag05} and to the design of crystallization solvents~\cite{kag06}.  Bommareddy et al.~\cite{bcse10} addressed the product/process design problem first in the space of the process, finding ranges of properties for the molecules to be designed.  These ranges then represented what was most suitable for a particular process and therefore defined a much smaller search space for the molecular design subproblem.

Bardow et al.~\cite{bsg10} proposed the CoMT-CAMD approach to first identify target solvent properties and then select an optimal solvent based on proximity to these ideal solvent properties.  This was applied in conjunction to a variant of the SAFT equation of state to design a carbon capture and storage process.  Stavrou et al.~\cite{slbg14} used the same approach to consider carbon capture problems.  Pereira et al.~\cite{pkg+11} also used SAFT to optimize a process to separate carbon dioxide and methane at high pressures.  Papadopoulos et al.~\cite{psl10} designed fluids for an organic Rankine cycle, considering fluids which fell on the Pareto front of optimal properties for the process.  Lampe et al. considered the same problem for fluid selection~\cite{lsb+14} and fluid design~\cite{lss+15}.  A summary of the references, categorized by application, is provided in Table~\ref{bigtable}.


\section{Conclusions}

The advent of the computational age has drastically impacted the design of chemical products and novel molecules, altering a once intuition-based, trial-and-error practice into a rapid and efficient search through millions of possible structures.  The availability and accuracy of QSPRs combined with efficient mathematical programming techniques has extended this capability even further, enabling chemical product designers to investigate a previously unimaginable diversity of chemical structures.

This article has provided background on the QSPRs which often serve as the underpinning of CAMD problems.  Each of three methods (group contribution, topological indices, and signature descriptors) was discussed in detail, and relevant constraints for optimization problems were provided in each case.  The CAMD problem was also addressed from the vantage point of mathematical optimization.  Various formulations were discussed for a few broad classes of the CAMD problem (single-molecule design, mixture design, integrated product/process design).  Solution techniques were discussed to aid in the solution of the often difficult CAMD problem.  Finally, we provided a summary of the many design endeavors and applications of the CAMD problem.

The increasing availability of computational resources, efficient optimization algorithms, and accurate QSPRs bodes well for the future of CAMD.  CAMD has a long history of proposing improved solutions for many well-known industrial processes as well as designing new products for consumers and optimizing high-impact chemical processes.  More recently, there have been a growing number of more exotic modeling and design efforts, concerning ideas such as integrating quantum chemistry techniques, designing transition metal catalysts, and determining optimal structures of pharmaceutical compounds.  The potential applications of CAMD are numerous, and the field is poised to play an integral role in the development of the chemical and biochemical technologies of the not-so-distant future.

\end{singlespacing}


\begin{thebibliography}{173}
\providecommand{\natexlab}[1]{#1}
\providecommand{\url}[1]{\texttt{#1}}
\expandafter\ifx\csname urlstyle\endcsname\relax
  \providecommand{\doi}[1]{doi: #1}\else
  \providecommand{\doi}{doi: \begingroup \urlstyle{rm}\Url}\fi

\bibitem[Abraham et~al.(1987)Abraham, Doherty, Kamlet, Harris, and
  Taft]{adk+87}
M.~H. Abraham, R.~M. Doherty, M.~J. Kamlet, J.~M. Harris, and R.~W. Taft.
\newblock {Linear solvation energy relationships. Part 37. An analysis of
  contributions of dipolarity--polarisability, nucleophilic assistance,
  electrophilic assistance, and cavity terms to solvent effects on t-butyl
  halide solvolysis rates}.
\newblock \emph{Journal of the Chemical Society, Perkin Transactions 2}, pages
  913--920, 1987.

\bibitem[Abrams and Prausnitz(1975)]{ap75}
D.~S. Abrams and J.~M. Prausnitz.
\newblock {Statistical thermodynamics of liquid mixtures: A new expression for
  the excess Gibbs energy of partly or completely miscible systems}.
\newblock \emph{AIChE Journal}, 21:\penalty0 116--128, 1975.

\bibitem[Austin et~al.(2016{\natexlab{a}})Austin, Sahinidis, and Trahan]{ast16}
N.~D. Austin, N.~V. Sahinidis, and D.~W. Trahan.
\newblock {A COSMO-based approach to computer-aided mixture design}.
\newblock \emph{Chemical Engineering Science}, 2016{\natexlab{a}}.
\newblock {DOI 10.1016/j.ces.2016.05.025}.

\bibitem[Austin et~al.(2016{\natexlab{b}})Austin, Samudra, Sahinidis, and
  Trahan]{asst16}
N.~D. Austin, A.~P. Samudra, N.~V. Sahinidis, and D.~W. Trahan.
\newblock Mixture design using derivative-free optimization in the space of
  individual component properties.
\newblock \emph{AIChE Journal}, 62:\penalty0 1514--1530, 2016{\natexlab{b}}.

\bibitem[Bajaj et~al.(2005)Bajaj, Sambi, and Madan]{bsm05}
S.~Bajaj, S.~S. Sambi, and A.~K. Madan.
\newblock {Prediction of anti-inflammatory activity of N-arylanthranilic acids:
  Computational approach using refined Zagreb indices}.
\newblock \emph{Croatica Chemica Acta}, 78:\penalty0 165--174, 2005.

\bibitem[Bardow et~al.(2010)Bardow, Steur, and Gross]{bsg10}
A.~Bardow, K.~Steur, and J.~Gross.
\newblock {Continuous-molecular targeting for integrated solvent and process
  design}.
\newblock \emph{Industrial \& Engineering Chemistry Research}, 49:\penalty0
  2834--2840, 2010.

\bibitem[Basak(1987)]{basak87}
S.~C. Basak.
\newblock {Use of molecular complexity indices in predictive pharmacology and
  toxicology: A QSAR approach}.
\newblock \emph{Medical Science Research}, 15:\penalty0 605--609, 1987.

\bibitem[Basak et~al.(1984)Basak, Gieschen, and Magnuson]{bgm84}
S.~C. Basak, D.~P. Gieschen, and V.~R. Magnuson.
\newblock {A quantitative correlation of the LC50 values of esters in
  Pimephales promelas using physicochemical and topological parameters}.
\newblock \emph{Environmental Toxicology and Chemistry}, 3:\penalty0 191--199,
  1984.

\bibitem[Benson(1999)]{b99}
S.~W. Benson.
\newblock New methods for estimating the heats of formation, heat capacities,
  and entropies of liquids and gases.
\newblock \emph{The Journal of Physical Chemistry A}, 103:\penalty0
  11481--11485, 1999.

\bibitem[Benson and Buss(1958)]{bb58}
S.~W. Benson and J.~H. Buss.
\newblock {Additivity rules for the estimation of molecular properties.
  Thermodynamic properties}.
\newblock \emph{The Journal of Chemical Physics}, 29:\penalty0 546--572, 1958.

\bibitem[Benson et~al.(1969)Benson, Cruickshank, Golden, Haugen, O'neal,
  Rodgers, Shaw, and Walsh]{bcg+69}
S.~W. Benson, F.~R. Cruickshank, D.~M. Golden, G.~R. Haugen, H.~E. O'neal,
  A.~S. Rodgers, R.~Shaw, and R.~Walsh.
\newblock Additivity rules for the estimation of thermochemical properties.
\newblock \emph{Chemical Reviews}, 69:\penalty0 279--324, 1969.

\bibitem[Boethling(1986)]{b86}
R.~S. Boethling.
\newblock Application of molecular topology to quantitative
  structure-biodegradability relationships.
\newblock \emph{Environmental Toxicology and Chemistry}, 5:\penalty0 797--806,
  1986.

\bibitem[Bohacek et~al.(1996)Bohacek, McMartin, and Guida]{bcg96}
R.~S. Bohacek, C.~McMartin, and W.~C. Guida.
\newblock {The art and practice of structure-based drug design: A molecular
  modeling perspective}.
\newblock \emph{Medicinal research reviews}, 16:\penalty0 3--50, 1996.

\bibitem[Bommareddy et~al.(2010)Bommareddy, Chemmangattuvalappil, Solvason, and
  Eden]{bcse10}
S.~Bommareddy, N.~G. Chemmangattuvalappil, C.~C. Solvason, and M.~R. Eden.
\newblock Simultaneous solution of process and molecular design problems using
  an algebraic approach.
\newblock \emph{Computers \& Chemical Engineering}, 34:\penalty0 1481--1486,
  2010.

\bibitem[Bonchev(1991)]{chemical_graph_theory}
D.~Bonchev.
\newblock \emph{Chemical graph theory: Introduction and fundamentals},
  volume~1.
\newblock CRC Press, 1991.

\bibitem[Brignole et~al.(1986)Brignole, Bottini, and Gani]{bbg86}
E.~A. Brignole, S.~B. Bottini, and R.~Gani.
\newblock {A strategy for the design and selection of solvents for separation
  processes}.
\newblock \emph{Fluid Phase Equilibria}, 29:\penalty0 125--132, 1986.

\bibitem[Brown et~al.(2006)Brown, Martin, Rintoul, and Faulon]{bmrf06}
W.~M. Brown, S.~Martin, M.~D. Rintoul, and J.-L. Faulon.
\newblock Designing novel polymers with targeted properties using the signature
  molecular descriptor.
\newblock \emph{Journal of chemical information and modeling}, 46:\penalty0
  826--835, 2006.

\bibitem[Burger et~al.(2015)Burger, Papaioannou, Gopinath, Jackson, Galindo,
  and Adjiman]{bpg+15}
J.~Burger, V.~Papaioannou, S.~Gopinath, G.~Jackson, A.~Galindo, and C.~S.
  Adjiman.
\newblock {A hierarchical method to integrated solvent and process design of
  physical CO2 absorption using the SAFT-$\gamma$ Mie approach}.
\newblock \emph{AIChE Journal}, 61:\penalty0 3249--3269, 2015.

\bibitem[Buxton et~al.(1999)Buxton, Livingston, and Pistikopoulos]{blp99}
A.~Buxton, A.~G. Livingston, and E.~N. Pistikopoulos.
\newblock {Optimal design of solvent blends for environmental impact
  minimization}.
\newblock \emph{AIChE Journal}, 45:\penalty0 817--843, 1999.

\bibitem[Camarda and Maranas(1999)]{cm99}
K.~V. Camarda and C.~D. Maranas.
\newblock {Optimization in polymer design using connectivity indices}.
\newblock \emph{Industrial \& Engineering Chemistry Research}, 38:\penalty0
  1884--1892, 1999.

\bibitem[Camarda and Sunderesan(2005)]{cs05}
K.~V. Camarda and P.~Sunderesan.
\newblock An optimization approach to the design of value-added soybean oil
  products.
\newblock \emph{Industrial \& Engineering Chemistry Research}, 44:\penalty0
  4361--4367, 2005.

\bibitem[Cao et~al.(1993)Cao, Knudsen, Fredenslund, and Rasmussen]{ckf+93}
W.~Cao, K.~Knudsen, A.~Fredenslund, and P.~Rasmussen.
\newblock {Group-contribution viscosity predictions of liquid mixtures using
  UNIFAC-VLE parameters}.
\newblock \emph{Industrial \& engineering chemistry research}, 32:\penalty0
  2088--2092, 1993.

\bibitem[Ceriani et~al.(2007)Ceriani, Gon{\c{c}}alves, Rabelo, Caruso, Cunha,
  Cavaleri, Batista, and Meirelles]{cgr+97}
R.~Ceriani, C.~B. Gon{\c{c}}alves, J.~Rabelo, M.~Caruso, A.~C.~C. Cunha, F.~W.
  Cavaleri, E.~A.~C. Batista, and A.~J.~A. Meirelles.
\newblock Group contribution model for predicting viscosity of fatty compounds.
\newblock \emph{Journal of Chemical \& Engineering Data}, 52:\penalty0
  965--972, 2007.

\bibitem[Ceriani et~al.(2009)Ceriani, Gani, and Meirelles]{cgm09}
R.~Ceriani, R.~Gani, and A.~J.~A. Meirelles.
\newblock Prediction of heat capacities and heats of vaporization of organic
  liquids by group contribution methods.
\newblock \emph{Fluid Phase Equilibria}, 283:\penalty0 49--55, 2009.

\bibitem[Chapman et~al.(1989)Chapman, Gubbins, Jackson, and Radosz]{saft}
W.~G. Chapman, K.~E. Gubbins, G.~Jackson, and M.~Radosz.
\newblock {SAFT: Equation-of-state solution model for associating fluids}.
\newblock \emph{Fluid Phase Equilibria}, 52:\penalty0 31--38, 1989.

\bibitem[Chavali et~al.(2004)Chavali, Lin, Miller, and Camarda]{clmc04}
S.~Chavali, B.~Lin, D.~C. Miller, and K.~V. Camarda.
\newblock Environmentally-benign transition metal catalyst design using
  optimization techniques.
\newblock \emph{Computers \& Chemical Engineering}, 28:\penalty0 605--611,
  2004.

\bibitem[{Chemical~Society}()]{cas}
American {Chemical~Society}.
\newblock {CAS Registry}.
\newblock Available at \url{www.cas.org/content/chemical-substances}.

\bibitem[Chemmangattuvalappil et~al.(2010)Chemmangattuvalappil, Solvason,
  Bommareddy, and Eden]{csbe10}
N.~G. Chemmangattuvalappil, C.~C. Solvason, S.~Bommareddy, and M.~R. Eden.
\newblock Reverse problem formulation approach to molecular design using
  property operators based on signature descriptors.
\newblock \emph{Computers \& Chemical Engineering}, 34:\penalty0 2062--2071,
  2010.

\bibitem[Churchwell et~al.(2004)Churchwell, Rintoul, Martin, Visco, Kotu,
  Larson, Sillerud, Brown, and Faulon]{crmvklsbf04}
C.~J. Churchwell, M.~D. Rintoul, S.~Martin, D.~P. Visco, A.~Kotu, R.~S. Larson,
  L.~O. Sillerud, D.~C. Brown, and J.-L. Faulon.
\newblock {The signature molecular descriptor: 3. Inverse-quantitative
  structure--activity relationship of ICAM-1 inhibitory peptides}.
\newblock \emph{Journal of Molecular Graphics and Modelling}, 22:\penalty0
  263--273, 2004.

\bibitem[Churi and Achenie(1996)]{ca96}
N.~Churi and L.~E.~K. Achenie.
\newblock Novel mathematical programming model for computer aided molecular
  design.
\newblock \emph{Industrial \& Engineering Chemistry Research}, 35:\penalty0
  3788--3794, 1996.

\bibitem[Cohen and Benson(1993)]{cb93}
N.~Cohen and S.~W. Benson.
\newblock Estimation of heats of formation of organic compounds by additivity
  methods.
\newblock \emph{Chemical Reviews}, 93:\penalty0 2419--2438, 1993.

\bibitem[Constantinou and Gani(1994)]{cg94}
L.~Constantinou and R.~Gani.
\newblock {New group contribution method for estimating properties of pure
  compounds}.
\newblock \emph{AIChE Journal}, 40:\penalty0 1697--1710, 1994.

\bibitem[Conte et~al.(2011)Conte, Gani, and Ng]{cgn11}
E.~Conte, R.~Gani, and K.~M. Ng.
\newblock {Design of formulated products: A systematic methodology}.
\newblock \emph{AIChE Journal}, 57:\penalty0 2431--2449, 2011.

\bibitem[Cussler and Moggridge(2011)]{cm11}
E.~L. Cussler and G.~D. Moggridge.
\newblock \emph{Chemical Product Design}.
\newblock Cambridge University Press, 2nd edition, 2011.

\bibitem[Devillers and Balaban(2000)]{top_index_book}
J.~Devillers and A.~T. Balaban.
\newblock \emph{Topological indices and related descriptors in QSAR and QSPAR}.
\newblock CRC Press, 2000.

\bibitem[Domalski and Hearing(1988)]{dh88}
E.~S. Domalski and E.~D. Hearing.
\newblock {Estimation of the thermodynamic properties of hydrocarbons at 298.15
  K}.
\newblock \emph{Journal of Physical and Chemical Reference Data}, 17:\penalty0
  1637--1678, 1988.

\bibitem[Duran and Grossmann(1986)]{dg86}
M.~A. Duran and I.~E. Grossmann.
\newblock {An outer-approximation algorithm for a class of mixed-integer
  nonlinear programs}.
\newblock \emph{Mathematical Programming}, 36:\penalty0 307--339, 1986.

\bibitem[Duvedi and Achenie(1996)]{da96}
A.~P. Duvedi and L.~E.~K. Achenie.
\newblock {Designing environmentally safe refrigerants using mathematical
  programming}.
\newblock \emph{Chemical Engineering Science}, 51:\penalty0 3727--3739, 1996.

\bibitem[Duvedi and Achenie(1997)]{da97}
A.~P. Duvedi and L.~E.~K. Achenie.
\newblock {On the design of environmentally benign refrigerant mixtures. A
  mathematical programming approach}.
\newblock \emph{Computers \& Chemical Engineering}, 21:\penalty0 915--923,
  1997.

\bibitem[Dyk and Nieuwoudt(2000)]{vn00}
B.~Van Dyk and I.~Nieuwoudt.
\newblock Design of solvents for extractive distillation.
\newblock \emph{Industrial \& Engineering Chemistry Research}, 39:\penalty0
  1423--1429, 2000.

\bibitem[Eden et~al.(2004)Eden, J{\o}rgensen, Gani, and El-Halwagi]{ejge04}
M.~R. Eden, S.~B. J{\o}rgensen, R.~Gani, and M.~M. El-Halwagi.
\newblock A novel framework for simultaneous separation process and product
  design.
\newblock \emph{Chemical Engineering and Processing: Process Intensification},
  43:\penalty0 595--608, 2004.

\bibitem[Eslick et~al.(2009)Eslick, Ye, Park, Topp, Spencer, and
  Camarda]{eyptsc09}
J.~C. Eslick, Q.~Ye, J.~Park, E.~M. Topp, P.~Spencer, and K.~V. Camarda.
\newblock A computational molecular design framework for crosslinked polymer
  networks.
\newblock \emph{Computers \& Chemical Engineering}, 33:\penalty0 954--963,
  2009.

\bibitem[Estrada(1995)]{e95}
E.~Estrada.
\newblock Edge adjacency relationships and a novel topological index related to
  molecular volume.
\newblock \emph{Journal of Chemical Information and Computer Sciences},
  35:\penalty0 31--33, 1995.

\bibitem[Estrada and Rodr{\'\i}guez(1999)]{er99}
E.~Estrada and L.~Rodr{\'\i}guez.
\newblock {Edge-connectivity indices in QSPR/QSAR studies. 1. Comparison to
  other topological indices in QSPR studies}.
\newblock \emph{Journal of Chemical Information and Computer Sciences},
  39:\penalty0 1037--1041, 1999.

\bibitem[Estrada and Uriarte(2001)]{eu01}
E.~Estrada and E.~Uriarte.
\newblock Recent advances on the role of topological indices in drug discovery
  research.
\newblock \emph{Current Medicinal Chemistry}, 8:\penalty0 1573--1588, 2001.

\bibitem[Faulon et~al.(2003{\natexlab{a}})Faulon, Churchwell, and Visco]{fcv03}
J.-L. Faulon, C.~J. Churchwell, and D.~P. Visco.
\newblock {The signature molecular descriptor. 2. Enumerating molecules from
  their extended valence sequences}.
\newblock \emph{Journal of Chemical Information and Computer Sciences},
  43:\penalty0 721--734, 2003{\natexlab{a}}.

\bibitem[Faulon et~al.(2003{\natexlab{b}})Faulon, Visco, and Pophale]{fvp03}
J.-L. Faulon, D.~P. Visco, and R.~S. Pophale.
\newblock {The signature molecular descriptor. 1. Using extended valence
  sequences in QSAR and QSPR studies}.
\newblock \emph{Journal of Chemical Information and Computer Sciences},
  43:\penalty0 707--720, 2003{\natexlab{b}}.

\bibitem[Foli\'{c} et~al.(2007)Foli\'{c}, Adjiman, and Pistikopoulos]{fap07}
M.~Foli\'{c}, C.~S. Adjiman, and E.~N. Pistikopoulos.
\newblock {Design of solvents for optimal reaction rate constants}.
\newblock \emph{AIChE Journal}, 53:\penalty0 1240--1256, 2007.

\bibitem[Folic et~al.(2008)Folic, Adjiman, and Pistikopoulos]{fap08}
M.~Folic, C.~S. Adjiman, and E.~N. Pistikopoulos.
\newblock {Computer-aided solvent design for reactions: Maximizing product
  formation}.
\newblock \emph{Industrial \& Engineering Chemistry Research}, 47:\penalty0
  5190--5202, 2008.

\bibitem[Fredenslund et~al.(1975)Fredenslund, Jones, and Prausnitz]{fjp75}
A.~Fredenslund, R.~L. Jones, and J.~M. Prausnitz.
\newblock {Group-contribution estimates of activity coefficients in nonideal
  liquid mixtures}.
\newblock \emph{AIChE Journal}, 21:\penalty0 1086, 1975.

\bibitem[Galvez et~al.(1994)Galvez, Garcia, Salabert, and Soler]{ggss94}
J.~Galvez, R.~Garcia, M.~T. Salabert, and R.~Soler.
\newblock {Charge indexes. New topological descriptors}.
\newblock \emph{Journal of Chemical Information and Computer Sciences},
  34:\penalty0 520--525, 1994.

\bibitem[Gani(2004)]{gani_camd_review}
R.~Gani.
\newblock Computer-aided methods and tools for chemical product design.
\newblock \emph{Chemical Engineering Research and Design}, 82:\penalty0
  1494--1504, 2004.

\bibitem[Gani and Brignole(1983)]{gb83}
R.~Gani and E.~A. Brignole.
\newblock {Molecular design of solvents for liquid extraction based on UNIFAC}.
\newblock \emph{Fluid Phase Equilibria}, 13:\penalty0 331--340, 1983.

\bibitem[Gani and Fredenslund(1993)]{gf93}
R.~Gani and A.~Fredenslund.
\newblock {Computer-aided molecular and mixture design with specified
  constraints}.
\newblock \emph{Fluid Phase Equilibria}, 82:\penalty0 39--46, 1993.

\bibitem[Gani et~al.(1991)Gani, Nielsen, and Fredenslund]{gnf91}
R.~Gani, B.~Nielsen, and A.~Fredenslund.
\newblock A group contribution approach to computer-aided molecular design.
\newblock \emph{AIChE Journal}, 37:\penalty0 1318--1332, 1991.

\bibitem[Gani et~al.(2005{\natexlab{a}})Gani, Harper, and Hostrup]{ghh05}
R.~Gani, P.~M. Harper, and M.~Hostrup.
\newblock Automatic creation of missing groups through connectivity index for
  pure-component property prediction.
\newblock \emph{Industrial \& Engineering Chemistry Research}, 44:\penalty0
  7262--7269, 2005{\natexlab{a}}.

\bibitem[Gani et~al.(2005{\natexlab{b}})Gani, Jim{\'e}nez-Gonz{\'a}lez, and
  Constable]{gjc05}
R.~Gani, C.~Jim{\'e}nez-Gonz{\'a}lez, and D.~J.~C. Constable.
\newblock Method for selection of solvents for promotion of organic reactions.
\newblock \emph{Computers \& Chemical Engineering}, 29:\penalty0 1661--1676,
  2005{\natexlab{b}}.

\bibitem[Gebreslassie and Diwekar(2015)]{gd15}
B.~H. Gebreslassie and U.~M. Diwekar.
\newblock {Efficient ant colony optimization for computer aided molecular
  design: Case study solvent selection problem}.
\newblock \emph{Computers \& Chemical Engineering}, 78:\penalty0 1--9, 2015.

\bibitem[Gharagheizi et~al.(2012)Gharagheizi, Gohar, and Vayeghan]{ggv12}
F.~Gharagheizi, M.~R.~S. Gohar, and M.~G. Vayeghan.
\newblock A quantitative structure--property relationship for determination of
  enthalpy of fusion of pure compounds.
\newblock \emph{Journal of Journal of thermal analysis and calorimetry},
  109:\penalty0 501--506, 2012.

\bibitem[Gopinath et~al.(2016)Gopinath, Jackson, Galindo, and Adjiman]{gjga16}
S.~Gopinath, G.~Jackson, A.~Galindo, and C.~S. Adjiman.
\newblock Outer approximation algorithm with physical domain reduction for
  computer-aided molecular and separation process design.
\newblock \emph{AIChE Journal}, 2016.

\bibitem[Gupta et~al.(2002)Gupta, Singh, and Madan]{gsm02}
S.~Gupta, M.~Singh, and A.~K. Madan.
\newblock {Application of graph theory: Relationship of eccentric connectivity
  index and Wiener's index with anti-inflammatory activity}.
\newblock \emph{Journal of Mathematical Analysis and Applications},
  266:\penalty0 259--268, 2002.

\bibitem[Hada et~al.(2014)Hada, Solvason, and Eden]{hse14}
S.~Hada, C.~C. Solvason, and M.~R. Eden.
\newblock Characterization-based molecular design of bio-fuel additives using
  chemometric and property clustering techniques.
\newblock \emph{Frontiers in Energy Research}, 2:\penalty0 20, 2014.

\bibitem[Hall and Kier(2007)]{hk07}
L.~H. Hall and L.~B. Kier.
\newblock The molecular connectivity chi indexes and kappa shape indexes in
  structure-property modeling.
\newblock \emph{Reviews in Computational Chemistry, Volume 2}, pages 367--422,
  2007.

\bibitem[Hall and Story(1996)]{hs96}
L.~H. Hall and C.~T. Story.
\newblock {Boiling point and critical temperature of a heterogeneous data set:
  QSAR with atom type electrotopological state indices using artificial neural
  networks}.
\newblock \emph{Journal of Chemical Information and Computer Sciences},
  36:\penalty0 1004--1014, 1996.

\bibitem[Hall et~al.(1975)Hall, Kier, and Murray]{hkm75}
L.~H. Hall, L.~B. Kier, and W.~J. Murray.
\newblock {Molecular connectivity II: Relationship to water solubility and
  boiling point}.
\newblock \emph{Journal of Pharmaceutical Sciences}, 64:\penalty0 1974--1977,
  1975.

\bibitem[Hall et~al.(1989{\natexlab{a}})Hall, Maynard, and Kier]{hmk89}
L.~H. Hall, E.~L. Maynard, and L.~B. Kier.
\newblock {QSAR investigation of benzene toxicity to fathead minnow using
  molecular connectivity}.
\newblock \emph{Environmental Toxicology and Chemistry}, 8:\penalty0 783--788,
  1989{\natexlab{a}}.

\bibitem[Hall et~al.(1989{\natexlab{b}})Hall, Maynard, and Kier]{hmk89_other}
L.~H. Hall, E.~L Maynard, and L.~B. Kier.
\newblock {Structure—activity relationship studies on the toxicity of benzene
  derivatives: III. Predictions and extension to new substituents}.
\newblock \emph{Environmental Toxicology and Chemistry}, 8:\penalty0 431--436,
  1989{\natexlab{b}}.

\bibitem[Hall et~al.(1993)Hall, Dailey, and Kier]{hdk93}
L.~H. Hall, R.~S. Dailey, and L.~B. Kier.
\newblock {Design of molecules from quantitative structure-activity
  relationship models. 3. Role of higher order path counts: Path 3}.
\newblock \emph{Journal of Chemical Information and Computer Sciences},
  33:\penalty0 598--603, 1993.

\bibitem[Harini et~al.(2013)Harini, Adhikari, and Rani]{rha13}
M.~Harini, J.~Adhikari, and K.~Y. Rani.
\newblock {A review of property estimation methods and computational schemes
  for rational solvent design: A focus on pharmaceuticals}.
\newblock \emph{Industrial \& Engineering Chemistry Research}, 52:\penalty0
  6869--6893, 2013.

\bibitem[Harper and Gani(2000)]{hg00}
P.~M. Harper and R.~Gani.
\newblock {A multi-step and multi-level approach for computer aided molecular
  design}.
\newblock \emph{Computers \& Chemical Engineering}, 24:\penalty0 677--683,
  2000.

\bibitem[Harper et~al.(1999)Harper, Gani, Kolar, and Ishikawa]{hg99}
P.~M. Harper, R.~Gani, P.~Kolar, and T.~Ishikawa.
\newblock {Computer-aided molecular design with combined molecular modeling and
  group contribution}.
\newblock \emph{Fluid Phase Equilibria}, 158--160:\penalty0 337--347, 1999.

\bibitem[Herring and Eden(2015)]{he15}
R.~H. Herring and M.~R. Eden.
\newblock Evolutionary algorithm for de novo molecular design with
  multi-dimensional constraints.
\newblock \emph{Computers \& Chemical Engineering}, 83:\penalty0 267--277,
  2015.

\bibitem[Hosoya(1971)]{h71}
H.~Hosoya.
\newblock {Topological index. A newly proposed quantity characterizing the
  topological nature of structural isomers of saturated hydrocarbons}.
\newblock \emph{Bulletin of the Chemical Society of Japan}, 44:\penalty0
  2332--2339, 1971.

\bibitem[Hosoya et~al.(1975)Hosoya, Hosoi, and Gutman]{hhg75}
H.~Hosoya, K.~Hosoi, and I.~Gutman.
\newblock A topological index for the total$\pi$-electron energy.
\newblock \emph{Theoretica Chimica Acta}, 38:\penalty0 37--47, 1975.

\bibitem[Hostrup et~al.(1999)Hostrup, M., and Gani]{hhg99}
M.~Hostrup, P.~M., and R.~Gani.
\newblock {Design of environmentally benign processes: Integration of design
  and separation process synthesis}.
\newblock \emph{Computers \& Chemical Engineering}, 23:\penalty0 1395--1414,
  1999.

\bibitem[Jalowka and Daubert(1986)]{jd86}
J.~W. Jalowka and T.~E. Daubert.
\newblock Group contribution method to predict critical temperature and
  pressure of hydrocarbons.
\newblock \emph{Industrial \& Engineering Chemistry Process Design and
  Development}, 25:\penalty0 139--142, 1986.

\bibitem[Joback(1989)]{joback89}
K.~G. Joback.
\newblock \emph{{Designing molecules possessing desired physical property
  values}}.
\newblock PhD thesis, Department of Chemical Engineering, Massachusetts
  Institute of Technology, Cambridge, MA, 1989.

\bibitem[Joback and Reid(1987)]{jr87}
K.~G. Joback and R.~C. Reid.
\newblock {Estimation of pure-component properties from group contributions}.
\newblock \emph{Chemical Engineering Communications}, 57:\penalty0 233--243,
  1987.

\bibitem[Joback and Stephanopoulos(1990)]{js90}
K.~G. Joback and G.~Stephanopoulos.
\newblock {Designing molecules possessing desired physical property values}.
\newblock \emph{Proceedings of the 1989 Foundations of Computer-Aided Process
  Design Conference, Snowmass, CO, {\em Elsevier, Amsterdam}}, pages 195--230,
  1990.

\bibitem[Joback and Stephanopoulos(1995)]{js95}
K.~G. Joback and G.~Stephanopoulos.
\newblock {Searching spaces of discrete solutions: The design of molecules
  possessing desired physical properties}.
\newblock \emph{Advances in Chemical Engineering}, 21:\penalty0 257--311, 1995.

\bibitem[Karunanithi and Mehrkesh(2013)]{km13}
A.~Karunanithi and A.~Mehrkesh.
\newblock Computer-aided design of tailor-made ionic liquids.
\newblock \emph{AIChE Journal}, 59:\penalty0 4627--4640, 2013.

\bibitem[Karunanithi et~al.(2005)Karunanithi, Achenie, and Gani]{kag05}
A.~T. Karunanithi, L.~E.~K. Achenie, and R.~Gani.
\newblock A new decomposition-based computer-aided molecular/mixture design
  methodology for the design of optimal solvents and solvent mixtures.
\newblock \emph{Industrial \& Engineering Chemistry Research}, 44:\penalty0
  4785--4797, 2005.

\bibitem[Karunanithi et~al.(2006)Karunanithi, Achenie, and Gani]{kag06}
A.~T. Karunanithi, L.~E.~K. Achenie, and R.~Gani.
\newblock {A computer-aided molecular design framework for crystallization
  solvent design}.
\newblock \emph{Chemical Engineering Science}, 61:\penalty0 1247--1260, 2006.

\bibitem[Katritzky and Gordeeva(1993)]{kg93}
A.~R. Katritzky and E.~V. Gordeeva.
\newblock {Traditional topological indexes vs electronic, geometrical, and
  combined molecular descriptors in QSAR/QSPR research}.
\newblock \emph{Journal of Chemical Information and Computer Sciences},
  33:\penalty0 835--857, 1993.

\bibitem[Katritzky et~al.(1998)Katritzky, Wang, Sild, Tamm, and
  Karelson]{kwstk98}
A.~R. Katritzky, Y.~Wang, S.~Sild, T.~Tamm, and M.~Karelson.
\newblock {QSPR studies on vapor pressure, aqueous solubility, and the
  prediction of water-air partition coefficients}.
\newblock \emph{Journal of Chemical Information and Computer Sciences},
  38:\penalty0 720--725, 1998.

\bibitem[Kauffman and Jurs(2001)]{kj01}
G.~W. Kauffman and P.~C. Jurs.
\newblock Prediction of surface tension, viscosity, and thermal conductivity
  for common organic solvents using quantitative structure-property
  relationships.
\newblock \emph{Journal of Chemical Information and Computer Sciences},
  41:\penalty0 408--418, 2001.

\bibitem[Kier and Hall(1976)]{kh76}
L.~B. Kier and L.~H. Hall.
\newblock {Molecular connectivity VII: specific treatment of heteroatoms}.
\newblock \emph{Journal of Pharmaceutical Sciences}, 65:\penalty0 1806--1809,
  1976.

\bibitem[Kier and Hall(1993)]{kh93}
L.~B. Kier and L.~H. Hall.
\newblock {The generation of molecular structures from a graph-based QSAR
  Equation}.
\newblock \emph{Quantitative Structure-Activity Relationships}, 12:\penalty0
  383--388, 1993.

\bibitem[Kier et~al.(1975)Kier, Hall, Murray, and Randi\'{c}]{khmr75}
L.~B. Kier, L.~H. Hall, W.~J. Murray, and M.~Randi\'{c}.
\newblock {Molecular connectivity I: Relationship to nonspecific local
  anesthesia}.
\newblock \emph{Journal of Pharmaceutical Sciences}, 64:\penalty0 1971--1974,
  1975.

\bibitem[Kier et~al.(1976)Kier, Murray, Randi{\'c}, and Hall]{kmwrh76}
L.~B. Kier, W.~J. Murray, M.~Randi{\'c}, and L.~H. Hall.
\newblock {Molecular connectivity V: connectivity series concept applied to
  density}.
\newblock \emph{Journal of Pharmaceutical Sciences}, 65:\penalty0 1226--1230,
  1976.

\bibitem[Kier et~al.(1993{\natexlab{a}})Kier, Hall, and Dailey]{khd93}
L.~B. Kier, L.~H. Hall, and R.~S. Dailey.
\newblock {Design of molecules from quantitative structure-activity
  relationship models. 3. Role of higher order path counts: Path 3}.
\newblock \emph{Journal of Chemical Information and Computer Sciences},
  33:\penalty0 598--603, 1993{\natexlab{a}}.

\bibitem[Kier et~al.(1993{\natexlab{b}})Kier, Hall, and Frazer]{khf93}
L.~B. Kier, L.~H. Hall, and J.~W. Frazer.
\newblock {Design of molecules from quantitative structure-activity
  relationship models 1. Information transfer between path and vertex degree
  counts}.
\newblock \emph{Journal of Chemical Information and Computer Sciences},
  33:\penalty0 143--147, 1993{\natexlab{b}}.

\bibitem[Kim and Diwekar(2002{\natexlab{a}})]{kd02}
K.~Kim and U.~M. Diwekar.
\newblock {Efficient combinatorial optimization under uncertainty. 2.
  Application to stochastic solvent selection}.
\newblock \emph{Industrial \& Engineering Chemistry Research}, 41:\penalty0
  1285--1296, 2002{\natexlab{a}}.

\bibitem[Kim and Diwekar(2002{\natexlab{b}})]{kd02b}
K.-J. Kim and U.~M. Diwekar.
\newblock Integrated solvent selection and recycling for continuous processes.
\newblock \emph{Industrial \& Engineering Chemistry Research}, 41:\penalty0
  4479--4488, 2002{\natexlab{b}}.

\bibitem[Klamt(1995)]{k95}
A.~Klamt.
\newblock {Conductor-like screening model for real solvents: A new approach to
  the quantitative calculation of solvation phenomena}.
\newblock \emph{The Journal of Physical Chemistry A}, 99:\penalty0 2224--2235,
  1995.

\bibitem[Klamt and Sch{\"u}{\"u}rmann(1993)]{ks93}
A.~Klamt and G.~Sch{\"u}{\"u}rmann.
\newblock {COSMO: A new approach to dielectric screening in solvents with
  explicit expressions for the screening energy and its gradient}.
\newblock \emph{Journal of the Chemical Society, Perkin Transactions},
  2:\penalty0 799--805, 1993.

\bibitem[Klamt et~al.(1998)Klamt, Jonas, B{\"u}rger, and Lohrenz]{kvbl98}
A.~Klamt, V.~Jonas, T.~B{\"u}rger, and J.~C.~W. Lohrenz.
\newblock {Refinement and parametrization of COSMO-RS}.
\newblock \emph{The Journal of Physical Chemistry A}, 102:\penalty0 5074--5085,
  1998.

\bibitem[Klein et~al.(1992)Klein, Wu, and Gani]{kwg92}
J.~A. Klein, D.~T. Wu, and R.~Gani.
\newblock Computer aided mixture design with specified property constraints.
\newblock \emph{Computers \& Chemical Engineering}, 16:\penalty0 S229--S236,
  1992.

\bibitem[Klincewicz and Reid(1984)]{kr84}
K.~M. Klincewicz and R.~C. Reid.
\newblock Estimation of critical properties with group contribution methods.
\newblock \emph{AIChE Journal}, 30:\penalty0 137--142, 1984.

\bibitem[Klopman and Zhu(2001)]{kz01}
G.~Klopman and H.~Zhu.
\newblock Estimation of the aqueous solubility of organic molecules by the
  group contribution approach.
\newblock \emph{Journal of Chemical Information and Computer Sciences},
  41:\penalty0 439--445, 2001.

\bibitem[Klopman et~al.(1994)Klopman, Li, Wang, and Dimayuga]{klwd94}
G.~Klopman, J.-Y. Li, S.~Wang, and M.~Dimayuga.
\newblock {Computer automated log P calculations based on an extended group
  contribution approach}.
\newblock \emph{Journal of Chemical Information and Computer Sciences},
  34:\penalty0 752--781, 1994.

\bibitem[Knight and McRae(1991)]{km91}
J.~P. Knight and G.~J. McRae.
\newblock {A combinatorial optimization approach to molecular design}.
\newblock \emph{Nanotechnology}, 2:\penalty0 142--148, 1991.

\bibitem[Kolsk{\'a} et~al.(2008)Kolsk{\'a}, Kukal, Z{\'a}bransk{\`y}, and
  Ruzˇicˇka]{kkzr08}
Z.~Kolsk{\'a}, J.~Kukal, M.~Z{\'a}bransk{\`y}, and V.~Ruzˇicˇka.
\newblock Estimation of the heat capacity of organic liquids as a function of
  temperature by a three-level group contribution method.
\newblock \emph{Industrial \& Engineering Chemistry Research}, 47:\penalty0
  2075--2085, 2008.

\bibitem[Lampe et~al.(2014)Lampe, Stavrou, Bücker, Gross, and Bardow]{lsb+14}
M.~Lampe, M.~Stavrou, H.~M. Bücker, J.~Gross, and A.~Bardow.
\newblock {Simultaneous optimization of working fluid and process for organic
  Rankine cycles using PC-SAFT}.
\newblock \emph{Industrial \& Engineering Chemistry Research}, 53:\penalty0
  8821--8830, 2014.

\bibitem[Lampe et~al.(2015)Lampe, Stavrou, Schilling, Sauer, Gross, and
  Bardow]{lss+15}
M.~Lampe, M.~Stavrou, J.~Schilling, E.~Sauer, J.~Gross, and A.~Bardow.
\newblock {Computer-aided molecular design in the continuous-molecular
  targeting framework using group-contribution PC-SAFT}.
\newblock \emph{Computers \& Chemical Engineering}, 81:\penalty0 278--287,
  2015.

\bibitem[Lin et~al.(2005)Lin, Chavali, Camarda, and Miller]{lc05}
B.~Lin, S.~Chavali, K.~Camarda, and D.~C. Miller.
\newblock {Computer-aided molecular design using tabu search}.
\newblock \emph{Computers \& Chemical Engineering}, 29:\penalty0 337--347,
  2005.

\bibitem[Lin and Sandler(2002)]{ls02iecr}
S.-T. Lin and S.~I. Sandler.
\newblock A priori phase equilibrium prediction from a segment contribution
  solvation model.
\newblock \emph{Industrial \& Engineering Chemistry Research}, 41:\penalty0
  899--913, 2002.

\bibitem[Lymperiadis et~al.(2007)Lymperiadis, Adjiman, Galindo, and
  Jackson]{lagj07}
A.~Lymperiadis, C.~S. Adjiman, A.~Galindo, and G.~Jackson.
\newblock {A group contribution method for associating chain molecules based on
  the statistical associating fluid theory (SAFT-$\gamma$)}.
\newblock \emph{The Journal of Chemical Physics}, 127:\penalty0 234903, 2007.

\bibitem[Lymperiadis et~al.(2008)Lymperiadis, Adjiman, Jackson, and
  Galindo]{lajg08}
A.~Lymperiadis, C.~S. Adjiman, G.~Jackson, and A.~Galindo.
\newblock A generalisation of the saft-group contribution method for groups
  comprising multiple spherical segments.
\newblock \emph{Fluid Phase Equilibria}, 274:\penalty0 85--104, 2008.

\bibitem[Macchietto et~al.(1990)Macchietto, Odele, and Omatsone]{moo90}
S.~Macchietto, O.~Odele, and O.~Omatsone.
\newblock {Design on optimal solvents for liquid-liquid extraction and gas
  absorption processes}.
\newblock \emph{Chemical Engineering Research and Design}, 68:\penalty0
  429--433, 1990.

\bibitem[Maranas(1996)]{mar96}
C.~D. Maranas.
\newblock {Optimal computer-aided molecular design: A polymer design case
  study}.
\newblock \emph{Industrial \& Engineering Chemistry Research}, 35:\penalty0
  3403--3414, 1996.

\bibitem[Marcoulaki and Kokossis(1998)]{mk98}
E.~C. Marcoulaki and A.~C. Kokossis.
\newblock {Molecular design synthesis using stochastic optimisation as a tool
  for scoping and screening}.
\newblock \emph{Computers \& Chemical Engineering}, 22:\penalty0 S11--S18,
  1998.

\bibitem[Marrero and Gani(2001)]{mg01}
J.~Marrero and R.~Gani.
\newblock {Group-contribution based estimation of pure component properties}.
\newblock \emph{Fluid Phase Equilibria}, 183--184:\penalty0 183--208, 2001.

\bibitem[Marrero and Gani(2002)]{mg02}
J.~Marrero and R.~Gani.
\newblock {Group-contribution based estimation of octanol/water partition
  coefficient and aqueous stability}.
\newblock \emph{Industrial \& Engineering Chemistry Research}, 41:\penalty0
  6623--6633, 2002.

\bibitem[Martin and Young(2001)]{my01}
T.~M. Martin and D.~M. Young.
\newblock {Prediction of the acute toxicity (96-h LC50) of organic compounds to
  the fathead minnow (Pimephales promelas) using a group contribution method}.
\newblock \emph{Chemical Research in Toxicology}, 14:\penalty0 1378--1385,
  2001.

\bibitem[Matsuda et~al.(2007)Matsuda, Yamamoto, Kurihara, and Tochigi]{mykt07}
H.~Matsuda, H.~Yamamoto, K.~Kurihara, and K.~Tochigi.
\newblock {Computer-aided reverse design for ionic liquids by QSPR using
  descriptors of group contribution type for ionic conductivities and
  viscosities}.
\newblock \emph{Fluid Phase Equilibria}, 261:\penalty0 434--443, 2007.

\bibitem[McLeese et~al.(2010)McLeese, Eslick, Hoffmann, Scurto, and
  Camarda]{mehsc10}
S.~E. McLeese, J.~C. Eslick, N.~J. Hoffmann, A.~M. Scurto, and K.~V. Camarda.
\newblock Design of ionic liquids via computational molecular design.
\newblock \emph{Computers \& Chemical Engineering}, 34:\penalty0 1476--1480,
  2010.

\bibitem[Mercader et~al.(2000)Mercader, Castro, and Toropov]{mct00}
A.~Mercader, E.~A. Castro, and A.~A. Toropov.
\newblock {QSPR modeling of the enthalpy of formation from elements by means of
  correlation weighting of local invariants of atomic orbital molecular
  graphs}.
\newblock \emph{Chemical Physics Letters}, 330:\penalty0 612--623, 2000.

\bibitem[Mu et~al.(2007)Mu, Rarey, and Gmehling]{mrg07}
T.~Mu, J.~Rarey, and J.~Gmehling.
\newblock {Group contribution prediction of surface charge density profiles for
  COSMO-RS(Ol)}.
\newblock \emph{AIChE Journal}, 53:\penalty0 3231--3240, 2007.

\bibitem[Mu et~al.(2009)Mu, Rarey, and Gmehling]{mrg09}
T.~Mu, J.~Rarey, and J.~Gmehling.
\newblock {Group contribution prediction of surface charge density distribution
  of molecules for COSMO-SAC}.
\newblock \emph{AIChE Journal}, 55:\penalty0 3298--3300, 2009.

\bibitem[Murray et~al.(1975)Murray, Hall, and Kier]{mhk75}
W.~J. Murray, L.~H. Hall, and L.~B. Kier.
\newblock {Molecular connectivity III: Relationship to partition coefficients}.
\newblock \emph{Journal of Pharmaceutical Sciences}, 64:\penalty0 1978--1981,
  1975.

\bibitem[Nannoolal et~al.(2004)Nannoolal, Rarey, Ramjugernath, and
  Cordes]{nrrc04}
Y.~Nannoolal, J.~Rarey, D.~Ramjugernath, and W.~Cordes.
\newblock {Estimation of pure component properties: Part 1. Estimation of the
  normal boiling point of non-electrolyte organic compounds via group
  contributions and group interactions}.
\newblock \emph{Fluid Phase Equilibria}, 226:\penalty0 45--63, 2004.

\bibitem[Nannoolal et~al.(2007)Nannoolal, Rarey, and Ramjugernath]{nrr07}
Y.~Nannoolal, J.~Rarey, and D.~Ramjugernath.
\newblock {Estimation of pure component properties: Part 2. Estimation of
  critical property data by group contribution}.
\newblock \emph{Fluid Phase Equilibria}, 252:\penalty0 1--27, 2007.

\bibitem[Nannoolal et~al.(2008)Nannoolal, Rarey, and Ramjugernath]{nrr08}
Y.~Nannoolal, J.~Rarey, and D.~Ramjugernath.
\newblock {Estimation of pure component properties: Part 3. Estimation of the
  vapor pressure of non-electrolyte organic compounds via group contributions
  and group interactions}.
\newblock \emph{Fluid Phase Equilibria}, 269:\penalty0 117--133, 2008.

\bibitem[Nannoolal et~al.(2009)Nannoolal, Rarey, and Ramjugernath]{nrr09}
Y.~Nannoolal, J.~Rarey, and D.~Ramjugernath.
\newblock {Estimation of pure component properties. Part 4: Estimation of the
  saturated liquid viscosity of non-electrolyte organic compounds via group
  contributions and group interactions}.
\newblock \emph{Fluid Phase Equilibria}, 281:\penalty0 97--119, 2009.

\bibitem[Ng et~al.(2015)Ng, Chong, and Chemmangattuvalappil]{ng_camd_review}
L.~Y. Ng, F.~K. Chong, and N.~G. Chemmangattuvalappil.
\newblock Challenges and opportunities in computer-aided molecular design.
\newblock \emph{Computers \& Chemical Engineering}, 81:\penalty0 115--129,
  2015.

\bibitem[Odele and Macchietto(1993)]{om93}
O.~Odele and S.~Macchietto.
\newblock {Computer aided molecular design: A novel method for optimal solvent
  selection}.
\newblock \emph{Fluid Phase Equilibria}, 82:\penalty0 47--54, 1993.

\bibitem[Ourique and Telles(1998)]{ot98}
J.~E. Ourique and A.~S. Telles.
\newblock Computer-aided molecular design with simulated annealing and
  molecular graphs.
\newblock \emph{Computers \& Chemical Engineering}, 22:\penalty0 S615--S618,
  1998.

\bibitem[Papadopoulos and Linke(2005)]{pl05}
A.~I. Papadopoulos and P.~Linke.
\newblock A unified framework for integrated process and molecular design.
\newblock \emph{Chemical Engineering Research and Design}, 83:\penalty0
  674--678, 2005.

\bibitem[Papadopoulos and Linke(2006{\natexlab{a}})]{pl06a}
A.~I. Papadopoulos and P.~Linke.
\newblock {Multiobjective molecular design for integrated process-solvent
  systems synthesis}.
\newblock \emph{AIChE Journal}, 52:\penalty0 1057--1070, 2006{\natexlab{a}}.

\bibitem[Papadopoulos and Linke(2006{\natexlab{b}})]{pl06b}
A.~I. Papadopoulos and P.~Linke.
\newblock Efficient integration of optimal solvent and process design using
  molecular clustering.
\newblock \emph{Chemical Engineering Science}, 61:\penalty0 6316--6336,
  2006{\natexlab{b}}.

\bibitem[Papadopoulos et~al.(2010)Papadopoulos, Stijepovic, and Linke]{psl10}
A.~I. Papadopoulos, M.~Stijepovic, and P.~Linke.
\newblock {On the systematic design and selection of optimal working fluids for
  organic Rankine cycles}.
\newblock \emph{Applied Thermal Engineering}, 30:\penalty0 760 --769, 2010.

\bibitem[Papaioannou et~al.(2014)Papaioannou, Lafitte, Avenda{\~n}o, Adjiman,
  Jackson, M{\"u}ller, and Galindo]{plaajg14}
V.~Papaioannou, T.~Lafitte, C.~Avenda{\~n}o, C.~S. Adjiman, G.~Jackson, E.~A.
  M{\"u}ller, and A.~Galindo.
\newblock {Group contribution methodology based on the statistical associating
  fluid theory for heteronuclear molecules formed from Mie segments}.
\newblock \emph{The Journal of chemical physics}, 140:\penalty0 054107, 2014.

\bibitem[Partington(1970)]{history_of_chemistry}
J.~R. Partington.
\newblock A history of chemistry.
\newblock 1970.

\bibitem[Patel et~al.(2009)Patel, Ng, and Mannan]{pnm09}
S.~J. Patel, D.~Ng, and M.~S. Mannan.
\newblock {QSPR flash point prediction of solvents using topological indices
  for application in computer aided molecular design}.
\newblock \emph{Industrial \& Engineering Chemistry Research}, 48:\penalty0
  7378--7387, 2009.

\bibitem[Pavurala and Achenie(2013)]{pa13}
N.~Pavurala and L.~E.~K. Achenie.
\newblock A mechanistic approach for modeling oral drug delivery.
\newblock \emph{Computers \& Chemical Engineering}, 57:\penalty0 196--206,
  2013.

\bibitem[Peng et~al.(2009)Peng, Goff, dos Ramos, and McCabe]{pgdm09}
Y.~Peng, K.~D. Goff, M.~C. dos Ramos, and C.~McCabe.
\newblock {Developing a predictive group-contribution-based SAFT-VR equation of
  state}.
\newblock \emph{Fluid Phase Equilibria}, 277:\penalty0 131--144, 2009.

\bibitem[Pereira et~al.(2011)Pereira, Keskes, Galindo, Jackson, and
  Adjiman]{pkg+11}
F.E. Pereira, E.~Keskes, A.~Galindo, G.~Jackson, and C.S. Adjiman.
\newblock {Integrated solvent and process design using a SAFT-VR thermodynamic
  description: High-pressure separation of carbon dioxide and methane}.
\newblock \emph{Computers \& Chemical Engineering}, 35:\penalty0 474--491,
  2011.

\bibitem[Pistikopoulos and Stefanis(1998)]{ps98}
E.~N. Pistikopoulos and S.~K. Stefanis.
\newblock {Optimal solvent design for environmental impact minimization}.
\newblock \emph{Computers \& Chemical Engineering}, 22:\penalty0 717--733,
  1998.

\bibitem[Platts et~al.(2000)Platts, Abraham, Butina, and Hersey]{pabh00}
J.~A. Platts, M.~H. Abraham, D.~Butina, and A.~Hersey.
\newblock {Estimation of molecular linear free energy relationship descriptors
  by a group contribution approach. 2. Prediction of partition coefficients}.
\newblock \emph{Journal of Chemical Information and Computer Sciences},
  40:\penalty0 71--80, 2000.

\bibitem[Raman and Maranas(1998)]{rm98}
V.~S. Raman and C.~D. Maranas.
\newblock {Optimization in product design with properties correlated with
  topological indices}.
\newblock \emph{Computers \& Chemical Engineering}, 22:\penalty0 747--763,
  1998.

\bibitem[Randic(1975)]{r75}
M.~Randic.
\newblock Characterization of molecular branching.
\newblock \emph{Journal of the American Chemical Society}, 97:\penalty0
  6609--6615, 1975.

\bibitem[Randic and Zupan(2001)]{rz01}
M.~Randic and J.~Zupan.
\newblock On interpretation of well-known topological indices.
\newblock \emph{Journal of Chemical Information and Computer Sciences},
  41:\penalty0 550--560, 2001.

\bibitem[Roganov et~al.(2005)Roganov, Pisarev, Emel'yanenko, and
  Verevkin]{rpe+05}
G.~N. Roganov, P.~N. Pisarev, V.~N. Emel'yanenko, and S.~P. Verevkin.
\newblock {Measurement and prediction of thermochemical properties. Improved
  Benson-type increments for the estimation of enthalpies of vaporization and
  standard enthalpies of formation of aliphatic alcohols}.
\newblock \emph{Journal of Chemical \& Engineering Data}, 50:\penalty0
  1114--1124, 2005.

\bibitem[Rose et~al.(2002)Rose, Hall, and Kier]{rhk02}
K.~Rose, L.~H. Hall, and L.~B. Kier.
\newblock Modeling blood-brain barrier partitioning using the
  electrotopological state.
\newblock \emph{Journal of Chemical Information and Computer Sciences},
  42:\penalty0 651--666, 2002.

\bibitem[Sahinidis et~al.(2003)Sahinidis, Tawarmalani, and Yu]{sty03}
N.~V. Sahinidis, M.~Tawarmalani, and M.~Yu.
\newblock {Design of alternative refrigerants via global optimization}.
\newblock \emph{AIChE Journal}, 49:\penalty0 1761--1775, 2003.

\bibitem[Samudra and Sahinidis(2013{\natexlab{a}})]{ss13}
A.~Samudra and N.~V. Sahinidis.
\newblock {Optimization-based framework for computer-aided molecular design}.
\newblock \emph{AIChE Journal}, 59:\penalty0 3686--3701, 2013{\natexlab{a}}.

\bibitem[Samudra and Sahinidis(2013{\natexlab{b}})]{ss13iecr}
A.~Samudra and N.~V. Sahinidis.
\newblock {Design of heat transfer media components for retail food
  refrigeration}.
\newblock \emph{Industrial \& Engineering Chemistry Research}, 52:\penalty0
  8518--8526, 2013{\natexlab{b}}.

\bibitem[Sastri and Rao(1992)]{sr92}
S.~R.~S. Sastri and K.~K. Rao.
\newblock A new group contribution method for predicting viscosity of organic
  liquids.
\newblock \emph{The Chemical Engineering Journal}, 50:\penalty0 9--25, 1992.

\bibitem[Scheffczyk et~al.(2016)Scheffczyk, Fleitmann, Schwarz, Lampe, Bardow,
  and Leonhard]{sfslb16}
J.~Scheffczyk, L.~Fleitmann, A.~Schwarz, M.~Lampe, A.~Bardow, and K.~Leonhard.
\newblock {COSMO-CAMD: A framework for optimization-based computer-aided
  molecular design using COSMO-RS}.
\newblock \emph{Chemical Engineering Science}, 2016.

\bibitem[Schramke et~al.(1999)Schramke, Murphy, Doucette, and Hintze]{smdwh99}
J.~A. Schramke, S.~F. Murphy, W.~J. Doucette, and W.~D. Hintze.
\newblock {Prediction of aqueous diffusion coefficients for organic compounds
  at 25 C}.
\newblock \emph{Chemosphere}, 38:\penalty0 2381--2406, 1999.

\bibitem[Siddhaye et~al.(2000)Siddhaye, Camarda, Topp, and Southard]{scts00}
S.~Siddhaye, K.~V. Camarda, E.~Topp, and M.~Southard.
\newblock {Design of novel pharmaceutical products via combinatorial
  optimization}.
\newblock \emph{Computers \& Chemical Engineering}, 24:\penalty0 701--704,
  2000.

\bibitem[Siddhaye et~al.(2004)Siddhaye, Camarda, Southard, and Topp]{scst04}
S.~Siddhaye, K.~Camarda, M.~Southard, and E.~Topp.
\newblock Pharmaceutical product design using combinatorial optimization.
\newblock \emph{Computers \& Chemical Engineering}, 28:\penalty0 425--434,
  2004.

\bibitem[Sinha et~al.(1999)Sinha, Achenie, and Ostrovsky]{sao99}
M.~Sinha, L.~E.~K. Achenie, and G.~M. Ostrovsky.
\newblock {Environmentally benign solvent design by global optimization}.
\newblock \emph{Computers \& Chemical Engineering}, 23:\penalty0 1381--1394,
  1999.

\bibitem[Sinha et~al.(2003)Sinha, Achenie, and Gani]{sag03}
M.~Sinha, L.~E.~K. Achenie, and R.~Gani.
\newblock Blanket wash solvent blend design using interval analysis.
\newblock \emph{Industrial \& Engineering Chemistry Research}, 42:\penalty0
  516--527, 2003.

\bibitem[Stavrou et~al.(2014)Stavrou, Lampe, Bardow, and Gross]{slbg14}
M.~Stavrou, M.~Lampe, A.~Bardow, and J.~Gross.
\newblock {Continuous molecular targeting--computer-aided molecular design
  (CoMT--CAMD) for simultaneous process and solvent design for CO2 capture}.
\newblock \emph{Industrial \& Engineering Chemistry Research}, 53:\penalty0
  18029--18041, 2014.

\bibitem[Stein and Brown(1994)]{sb94}
S.~E. Stein and R.~L. Brown.
\newblock Estimation of normal boiling points from group contributions.
\newblock \emph{Journal of Chemical Information and Computer Sciences},
  34:\penalty0 581--587, 1994.

\bibitem[Struebing et~al.(2013)Struebing, Ganase, Karamertzanis, Siougkrou,
  Haycock, Piccione, Armstrong, Galindo, and Adjiman]{sgk+13}
H.~Struebing, Z.~Ganase, P.~G. Karamertzanis, E.~Siougkrou, P.~Haycock, P.~M.
  Piccione, A.~Armstrong, A.~Galindo, and C.~S. Adjiman.
\newblock Computer-aided molecular design of solvents for accelerated reaction
  kinetics.
\newblock \emph{Nature Chemistry}, 5:\penalty0 952--957, 2013.

\bibitem[Tihic et~al.(2007)Tihic, Kontogeorgis, von Solms, Michelsen, and
  Constantinou]{tkvmc07}
A.~Tihic, G.~M. Kontogeorgis, N.~von Solms, M.~L. Michelsen, and
  L.~Constantinou.
\newblock {A predictive group-contribution simplified PC-SAFT equation of
  state: Application to polymer systems}.
\newblock \emph{Industrial \& Engineering Chemistry Research}, 47:\penalty0
  5092--5101, 2007.

\bibitem[Vaidyanathan and El-Halwagi(1996)]{re96}
R.~Vaidyanathan and M.~El-Halwagi.
\newblock Computer-aided synthesis of polymers and blends with target
  properties.
\newblock \emph{Industrial \& Engineering Chemistry Research}, 35:\penalty0
  627--634, 1996.

\bibitem[Venkatasubramanian et~al.(1994)Venkatasubramanian, Chan, and
  Caruthers]{vcc94}
V.~Venkatasubramanian, K.~Chan, and J.~M. Caruthers.
\newblock {Computer-aided molecular design using genetic algorithms}.
\newblock \emph{Computers \& Chemical Engineering}, 18:\penalty0 833--844,
  1994.

\bibitem[Venkatasubramanian et~al.(1995)Venkatasubramanian, Chan, and
  Caruthers]{vcc95}
V.~Venkatasubramanian, K.~Chan, and J.~M. Caruthers.
\newblock {Evolutionary design of molecules with desired properties using the
  genetic algorithm}.
\newblock \emph{Journal of Chemical Information and Computer Sciences},
  35:\penalty0 188--195, 1995.

\bibitem[Visco et~al.(2002)Visco, Pophale, Rintoul, and Faulon]{vprf02}
D.~P. Visco, R.~S. Pophale, M.~D. Rintoul, and J.-L. Faulon.
\newblock Developing a methodology for an inverse quantitative
  structure-activity relationship using the signature molecular descriptor.
\newblock \emph{Journal of Molecular Graphics and Modelling}, 20:\penalty0
  429--438, 2002.

\bibitem[Wang and Achenie(2002)]{wa02b}
Y.~Wang and L.~E.~K. Achenie.
\newblock {Computer aided solvent design for extractive fermentation}.
\newblock \emph{Fluid Phase Equilibria}, 201:\penalty0 1--18, 2002.

\bibitem[Weis and Visco(2010)]{wv10}
D.~C. Weis and D.~P. Visco.
\newblock {Computer-aided molecular design using the Signature molecular
  descriptor: Application to solvent selection}.
\newblock \emph{Computers \& Chemical Engineering}, 34:\penalty0 1018--1029,
  2010.

\bibitem[Wiener(1947)]{w47}
H.~Wiener.
\newblock Structural determination of paraffin boiling points.
\newblock \emph{Journal of the American Chemical Society}, 69:\penalty0 17--20,
  1947.

\bibitem[Wu et~al.(2012)Wu, Zhang, Goldberg, Cohen, Pan, Arpin, and
  Bar-Yosef]{wzg+12}
X.~Wu, C.~Zhang, P.~Goldberg, D.~Cohen, Y.~Pan, T.~Arpin, and O.~Bar-Yosef.
\newblock {Early pottery at 20,000 years ago in Xianrendong Cave, China}.
\newblock \emph{Science}, 336:\penalty0 1696--1700, 2012.

\bibitem[Xu and Diwekar(2005)]{xd05}
W.~Xu and U.~M. Diwekar.
\newblock {Improved genetic algorithms for deterministic optimization and
  optimization under uncertainty. Part II. Solvent selection under
  uncertainty}.
\newblock \emph{Industrial \& Engineering Chemistry Research}, 44:\penalty0
  7138--7146, 2005.

\bibitem[Yao et~al.(2003)Yao, Fan, Doucet, Panaye, Liu, Zhang, Zhang, and
  Hu]{yfdplzzh03}
X.~Yao, B.~Fan, J.~P. Doucet, A.~Panaye, M.~Liu, R.~Zhang, X.~Zhang, and Z.~Hu.
\newblock Quantitative structure property relationship models for the
  prediction of liquid heat capacity.
\newblock \emph{QSAR \& Combinatorial Science}, 22:\penalty0 29--48, 2003.

\bibitem[Zhang et~al.(2015)Zhang, Cignitti, and Gani]{zcg15}
L.~Zhang, S.~Cignitti, and R.~Gani.
\newblock Generic mathematical programming formulation and solution for
  computer-aided molecular design.
\newblock \emph{Computers \& Chemical Engineering}, 78:\penalty0 79--84, 2015.

\bibitem[Zhou et~al.(2015)Zhou, Lyu, Qi, and Sundmacher]{zlqs15}
T.~Zhou, Z.~Lyu, Z.~Qi, and K.~Sundmacher.
\newblock {Robust design of optimal solvents for chemical reactions--A combined
  experimental and computational strategy}.
\newblock \emph{Chemical Engineering Science}, 137:\penalty0 613--625, 2015.

\bibitem[Zhou et~al.(2016{\natexlab{a}})Zhou, Wang, McBride, and
  Sundmacher]{zwms16}
T.~Zhou, J.~Wang, K.~McBride, and K.~Sundmacher.
\newblock Optimal design of solvents for extractive reaction processes.
\newblock \emph{AIChE Journal}, 2016{\natexlab{a}}.

\bibitem[Zhou et~al.(2016{\natexlab{b}})Zhou, Zhou, and Sundmacher]{zzs16}
T.~Zhou, Y.~Zhou, and K.~Sundmacher.
\newblock A hybrid stochastic--deterministic optimization approach for
  integrated solvent and process design.
\newblock \emph{Chemical Engineering Science}, 2016{\natexlab{b}}.

\end{thebibliography}
\end{document}